\documentclass[useAMS,usenatbib,usegraphicx]{mn2e}
\usepackage{amsmath,amssymb}
\usepackage{url}
\usepackage{fixltx2e}
\usepackage[T1]{fontenc}
\usepackage{times}

\newcommand{\msun}{\ensuremath{\mathrm{M}_{\odot}}}
\begin{document}

\title[Magnetised Pillars and Globules]
{Effects of Magnetic Fields on Photoionised Pillars and Globules}
\author[J.\ Mackey and A.J.\ Lim]
{
  Jonathan Mackey$^{1,2}$\thanks{email: \texttt{JMackey@astro.uni-bonn.de}} and 
  Andrew J. Lim$^1$\thanks{email: \texttt{Andy@cp.dias.ie}}\\
  $^1$Dublin Institute for Advanced Studies, 31 Fitzwilliam Place,
  Dublin 2, Ireland.\\
  $^2$Argelander Institut f\"ur Astronomie, Auf dem H\"ugel 71, 53121
  Bonn, Germany.
}
\date{November 2010:
  The definitive version is available at \texttt{http://www.blackwell-synergy.com}
}
\maketitle

\begin{abstract}
  The effects of initially uniform magnetic fields on the formation
  and evolution of dense pillars and cometary globules at the
  boundaries of H~\textsc{II} regions are investigated using 3D
  radiation-magnetohydrodynamics simulations.  It is shown, in
  agreement with previous work, that a strong initial magnetic field
  is required to significantly alter the non-magnetised dynamics
  because the energy input from photoionisation is so large that it
  remains the dominant driver of the dynamics in most situations.
  Additionally it is found that for weak and medium field strengths an
  initially perpendicular field is swept into alignment with the
  pillar during its dynamical evolution, matching magnetic field
  observations of the `Pillars of Creation' in M16 and also some
  cometary globules.  A strong perpendicular magnetic field remains in
  its initial configuration and also confines the photoevaporation
  flow into a bar-shaped dense ionised ribbon which partially shields
  the ionisation front and would be readily observable in
  recombination lines.  A simple analytic model is presented to
  explain the properties of this bright linear structure.  These
  results show that magnetic field strengths in star-forming regions
  can in principle be significantly constrained by the morphology of
  structures which form at the borders of H~\textsc{II} regions.
\end{abstract}

\begin{keywords}
  methods: numerical -- MHD -- radiative transfer -- H~\textsc{II}
  Regions -- ISM: magnetic fields
\end{keywords}

% ***************************************************
\section{Introduction}
\label{sec:introduction}
% ***************************************************
The study of elephant trunks, pillars and globules, found at the
borders of H~\textsc{II} regions around massive stars, has received
significant attention in recent years, both from an observational
perspective and in theoretical and computational models.  The famous
`Pillars of Creation' in M16 were observed at optical wavelengths with
HST~\citep{HesScoSanEA96}, in the
IR~\citep{IndRobWhiEA07,SugWatTamEA07}, and in
sub-mm/radio~\citep[e.g.][]{Pou98,WhiNelHolEA99}, showing that these
are dynamic structures with ongoing star formation which may or may
not have been triggered by the radiation which has shaped the pillars.
\citet{SmiPovWhiEA10} provide strong evidence that pillars in the
Carina Nebula are significant sites of sequential star formation
propagating away from the older star clusters in this region, building
on previous observations of synchronised star formation around the
periphery of the nebula by~\citet{SmiBro07}.  On smaller scales,
studies of T-Tauri star ages in the Orion
Nebula~\citep{LeeCheZhaEA05,LeeChe07} show decreasing stellar ages
moving away from massive stars and towards bright-rimmed clouds at the
H~\textsc{II} region/molecular cloud interface, again strongly
suggesting at least sequential and possibly triggered star formation.
The clear relationship between pillars/globules and second generation
star formation around OB associations, and the question of the extent
to which this star formation is triggered, provides strong motivation
to understand the formation and evolution of these structures.

In a previous paper~\citep[][hereafter ML10]{MacLim10} we investigated
the formation and evolution of dense pillars of gas and dust --
elephant trunks -- on the boundaries of H~\textsc{II} regions using 3D
hydrodynamical simulations including photoionising radiative transfer
(R-HD). It was found that shadowing of ionising radiation by an
inhomogeneous density field naturally forms elephant trunks without
the assistance of self-gravity, or of ionisation front and cooling
instabilities.  A combination of radiation-driven
implosion~\citep[RDI;][]{Ber89} and acceleration due to the rocket
effect~\citep{OorSpi55} produce elongated structures: RDI compresses
neutral gas until pressure equilibrium with ionised gas is achieved;
the rocket effect accelerates gas away from the radiation source
producing dynamic elongated structures with lifetimes of a few hundred
kyr (depending on clump masses/densities).  Strong neutral gas cooling
was found to enhance this formation mechanism, producing denser and
longer lived pillar-like structures compared to models with weak
cooling.

Models such as these for the formation of bright-rimmed clouds,
globules and pillars have been considered for many
years~\citep[e.g.][]{Pot58, Mar70, BodTenYor79, SanWhiKle82}; much of
this work is summarised by~\citet{Yor86}.  The RDI of a photoionised
clump and its subsequent acceleration and evolution was calculated
analytically~\citep{Ber89,BerMcK90} and subsequently numerically
by~\citet{LeFLaz94}.  \citet{WilWarWhi01} considered a range of
axisymmetric models showing that multiple scenarios could form
long-lived pillar-like structures. 
It has been shown \citep{KesBur03, MiaWhiNelEA06, PouKanRyuEA07, BisWhiWueEA10} that RDI of
single clumps can generate cometary globules and trigger gravitational
collapse,
but it is more difficult to form pillars like those in M16
because gas must accumulate to a high density in the shadowed tail
region.  \citet{LimMel03} showed how photoionisation of multiple
clumps which partially shadow each other leads to dense gas
accumulating in shadowed tail regions; it was suggested
by~\citet{PouKanRyuEA07} that multiple clumps are required to
quantitatively match observations of the pillars in M16. Recent
models~\citep[][ML10]{MelArtHenEA06, RagHenVasEA09, LorRagEsq09,
  GriNaaWalEA09, GriBurNaaEA10} showed that elephant trunks can form
quite naturally from the photoionisation of a clumpy density field
under a range of initial conditions.  \citet{GriBurNaaEA10} extended
our previous work (ML10) which used static initial conditions by
considering the photoionisation of dynamic density fields generated by
isothermal decaying turbulence, measuring the formation of pillar-like
structures as a function of turbulent Mach number.

Filamentary structure in an apparently helical geometry was found in
the shadowed tail regions of a number of elephant trunks\
\citep{CarKriGah98, CarGahKri02, CarGahKri03}, and it was suggested
this could arise due to twisting of magnetic field lines.  Velocity
profiles for some trunks were shown to be consistent with solid-body
rotation~\citep{GahCarJohEA06}, again consistent with a magnetic
origin of the observed structure, although the actual magnetic field
orientation and strength has not yet been measured for these
structures.  The magnetic field in cometary globules has been measured
by optical polarimetry~\citep{SriBhaRaj96,Bha99,BhaMahMan04}, showing
in two cases a field orientation along the head--tail axis of the
globule, but in one case~\citep{Bha99} a perpendicular field was
found.  \citet{SugWatTamEA07} used near-IR polarimetry of background
stars to measure the magnetic field in M16, finding an ordered
large-scale field in the H~\textsc{II} region, but within the pillars
the field is aligned with the pillar axes and significantly misaligned
with the ambient field by $\theta \sim 30$--$40^{\circ}$.  They
suggest this should constrain the magnetic field strength since it has
not been strong enough to resist reorientation during the formation
and evolution of the pillars, a suggestion we explore in more detail
in this work.

Theoretical calculations of the effects of magnetic fields on the
expansion of H~\textsc{II} regions were first considered
by~\citet{Las66b}.  Using analytic calculations \citet{RedWilDysEA98}
studied ionisation fronts with a plane-parallel magnetic field in the
plane of the ionisation front. They found that the velocity separation
between R-type and D-type solutions decreases with increasing field
strength, and that D-critical ionisation fronts also advance into the
neutral gas more rapidly for increasing field strength.
In~\citet{WilDysHar00} jump conditions for ionisation fronts with
oblique magnetic fields were presented, together with 1D numerical
models showing how an ionisation front could progress from fast-R-type
through fast-D-type, slow-R-type, and finally to slow-D-type.  These
extra modes are allowed because fast and slow shocks detach from the
ionisation front at different times and propagate into the neutral
medium.  This work was extended by~\citet{WilDys01}, who calculated
the internal structures of stationary 1D ionisation fronts.  They
found that oblique fields could produce significant transverse
velocities and regions where the transverse field component is
significantly modified.

\citet{Wil07} used 2D slab-symmetric radiation-magnetohydrodynamics
(R-MHD) simulations to study the photoevaporation flows from
magnetised globules.  In these models clumps with an initial density
of $n_\mathrm{H}=2\times10^5\,\mathrm{cm}^{-3}$ were placed in an
ambient medium $10\times$ less dense.  For simplicity a uniform
magnetic field with various strengths and orientations was used.
Plane-parallel radiation was assumed, with a thermal model where the
temperature relaxed to a value between $100\,$K and $10\,000\,$K
according to its ionisation fraction.  It was found that for a weak
field the ionised gas pressure dominates the dynamics and the field
was swept into a configuration where it was parallel to the column of
neutral gas behind the dense clump.  For a sufficiently strong field,
however, the field determined the dynamics and made the flow almost
one-dimensional along field lines.

The first 3D R-MHD calculation including non-equilibrium photoionising
radiative transfer was performed by~\citet*{KruStoGar07}.  They used
the \textsc{Athena} code with a new ray-tracing module to simulate the
expansion of an H~\textsc{II} region around a point source in a uniform
magnetic field.  The overall expansion is now axisymmetric rather than
spherically symmetric~\citep[cf.][]{Las66a} with a dense shell forming
in directions parallel to the field, and a possibly numerical
instability developing for expansion perpendicular to the field.

Using similar methods,~\citet{HenArtDeCEA09} model the photoionisation
of a dense clump of gas in 3D with an initially uniform magnetic
field.  They use the photon-conserving C$^2$-ray ray-tracing
algorithm~\citep{MelIliAlvEA06} which allows large timesteps to be
taken without loss of accuracy.  They found that the evolution of a
photoionised globule can be significantly altered by the presence of a
strong field, and in some cases a recombining shell forms at the
termination shock of the photoevaporation flow when it is confined by
a transverse field.  The implosion phase is strongly asymmetric since
the clump compresses much more readily along field lines than across
them.  These authors introduce a detailed thermal model to model the
dynamics as realistically as possible for conditions in the Orion
Nebula, finding that X-ray heating keeps the neutral gas temperature
at $T\gtrsim50\,$K, but the cooling in neutral gas is somewhat
stronger than that considered in ML10.

In this work we add magnetic fields of various strengths and
orientations to some of the models considered in ML10 to study their
effects on the dynamics of the dense neutral gas.  The simulation code
is described in \S\ref{sec:numerics}: \S\ref{ssec:MHD} describes the
MHD dynamics algorithm; \S\ref{ssec:MP} reviews the microphysical
processes included; \S\ref{ssec:KSG07} presents a brief comparison of
results obtained with our code to the test problem described
by~\citet{KruStoGar07}.  The 3D simulations presented here are
introduced in \S\ref{sec:simulations} and our results presented in
\S\ref{sec:results}.  \S\ref{ssec:projections} shows the evolution of
the projected gas density and magnetic field orientation;
\S\ref{ssec:emission} shows the emission from ionised gas;
\S\ref{ssec:R8R8a} evaluates boundary effects in strong field
simulations.  A calculation explaining the presence of a bright
ionised linear ridge/ribbon in the strong field simulations is presented in
\S\ref{sec:BarFormation}.  These results are compared to observations
of H~\textsc{II} regions and their magnetic fields in
\S\ref{sec:discussion}, where we also discuss our work in the context
of other recent computational research.  Our conclusions are presented
in \S\ref{sec:conclusions}, and some technical details and tests for
the simulations are given in the appendices.

% ***************************************************
\section{Numerical Methods}
\label{sec:numerics}
% ***************************************************
\subsection{MHD algorithms and tests}
\label{ssec:MHD}
These calculations are performed on a uniform finite-volume grid, with
the hydrodynamics and ray-tracing/microphysics modules as previously
described in ML10.  The simulations presented here use the same code,
but with an MHD module coupled to the ray-tracing.  The integration
scheme was described by~\citet{FalKomJoa98} and is second order
accurate in time and space, using the symmetric van Albada slope
limiter to ensure monotonicity near shocks~\citep{vAlbvLeeRob82}.  To
calculate the MHD fluxes a number of Riemann solvers were tried; we
use the Roe solver in conserved variables described
by~\citet{CarGal97} as it was the most robust for these calculations
(with corrections to typos obtained by comparison
with~\citealt{StoGarTeuEA08}).  Magnetic field units are used in the
code such that factors of $4\pi$ do not appear in the dynamical
equations~\citep[see e.g.][]{FalKomJoa98}; c.g.s.~units can be
obtained by scaling the field strength by $\sqrt{4\pi}$.  To avoid
confusion we will always quote field strength in c.g.s.\ units
(usually micro-Gauss, $\mu$G) rather than the code units.  We use
notation such that the total energy per unit volume is
$E=\frac{1}{2}\rho v^2 +p_g/(\gamma-1) +p_m$, being the sum of the
kinetic, internal (thermal), and magnetic energies.  The gas pressure
is $p_g$, magnetic pressure is $p_m=B^2/8\pi$, and the ideal gas
adiabatic index $\gamma$ is set to $5/3$.  Multidimensional
calculations use the artificial diffusion prescription
of~\citet{FalKomJoa98} to fix the `carbuncle problem' and to prevent
rarefaction shocks developing; typically a viscosity parameter
$\eta_v=0.1$ is used.

The code uses cell-centred values for vector quantities; the
divergence constraint in the magnetic field is addressed using the
`mixed-GLM' divergence cleaning method of~\citet{DedKemKroEA02}.
This algorithm introduces an extra scalar field which couples to
$\mathbf{\nabla}\cdot\mathbf{B}$, advecting and diffusing divergence
errors so that they do not build up near shocks or other
discontinuities.  While it does allow non-zero divergence of
$\mathbf{B}$ (which can introduce small force errors), the method has
some advantages.  (1) It is fully conservative in all of the physical
variables; only the evolution equation for the unphysical scalar
field, $\psi$, contains a source term.  (2) The total energy and
magnetic field updates are calculated in the same algorithmic step,
ensuring they are consistent.  An energy correction is still required,
but this can be calculated from the fluxes generated by divergence
cleaning.  This means the internal energy (as a fraction of the total
energy) is treated quite accurately, making the code robust for a wide
range of field strengths.  (3) It involves relatively little
computational and memory overhead.

A number of 1D and multidimensional adiabatic MHD code tests have been
performed to validate the code and for comparison to other
astrophysical MHD codes. Results from a number of these test problems
are shown in Appendix~\ref{app:MHDtests} and can also be found in more
detail at \url{http://www.astro.uni-bonn.de/~jmackey/jmac/} together
with HD and photoionisation test problems (some of which were
presented in ML10). 

Three modifications have been made to the mixed-GLM divergence
cleaning method.  (1) The parameter $c_r$ is set to $c_r=4\Delta x$
(where the grid cell diameter is $\Delta x$) instead of the
recommended value of $c_r=0.18$~\citep[see][eq.~45]{DedKemKroEA02}.  A
typo in their paper (A.~Dedner, private communication) meant $c_r$
should have been defined as $c_r=\Delta x/0.18$, similar to the value
used here.  (2) The algorithm induces extra magnetic field transport
across cell boundaries; the energy associated with this is accounted
for by an extra term in the energy flux:
\begin{equation}
  F[E] \rightarrow F[E]+ F[B_x]B_{x,m} \;. 
\end{equation}
Here $F[X]$ represents the flux of conserved variable $X$ across a
cell boundary, $B_x$ is the component of $\mathbf{B}$ normal to the
cell boundary, and $B_{x,m}$ is the value of $B_x$ in the interface
state calculated using the mixed-GLM
algorithm~\citep[][eq.~42]{DedKemKroEA02}.  This correction removed
gas pressure dips which appeared ahead of oblique shocks in
axisymmetric models of magnetised jets. (3) Motivated by numerical
difficulties encountered for subsonic flow across boundaries in
magnetically dominated media, the boundary condition for the scalar
field $\psi$ in the mixed-GLM method is modified for zero-gradient
boundaries.  This is described in Appendix~\ref{app:PsiBC} showing the
improvement obtained with the new boundary condition.

\subsection{Coupling MHD to photoionisation}
\label{ssec:MP}
We briefly review the microphysics algorithm here; the reader is
referred to ML10 for further details.  The dynamics and microphysics
are solved separately in turn by operator splitting: in a given
timestep the adiabatic MHD equations are solved and updated first
(with the ion fraction advected with the gas flow), and then the
internal energy per unit mass ($e_{\mathrm{int}}$) and ion fraction of
Hydrogen ($x$) are updated by an adaptive sub-stepping integration to
a user-specified relative accuracy.  In the absence of photoionisation
this uses a 5th order Runge--Kutta integration.  Ray-tracing uses the
short characteristics method with the photon-conserving discretisation
of the photoionisation rate given by~\citet{MelIliAlvEA06}.  When a
cell is strongly photoionised explicit integration is very unstable
and so the `constant electron density' approximation
of~\citet{SchKoe87} is used (see also~\citealt{MelIliAlvEA06}) to
analytically integrate the ion fraction when $x>0.95$.  The
time-averaged fraction of photons which pass through the cell is
calculated simultaneously to obtain the appropriate cell optical depth
for calculating rays to more distant grid cells.  The algorithm used
in ML10 was designed to work in an identical way for R-MHD as for
R-HD. As such, the non-dynamical test problems presented in ML10
produce identical results when run with a non-zero magnetic field.

We continue to use the on-the-spot approximation for diffuse
radiation; this remains a potential limitation of our results which
should be checked with future work. Recent calculations including
diffuse radiation~\citep{RagHenVasEA09,WilHen09} have shown that its
role in the dynamics of H~\textsc{II} regions may not be as significant as was
suggested by e.g.~\citet{Rit05}. We plan to test this in dynamical
simulations using an algorithm such as that described
by~\citet{KuiKlaDulEA10}; it should be possible to extend their method
to treat ionising photons.

Radiative cooling processes are modelled using the C2 cooling function
from ML10 -- in photoionised gas the coolants considered are
collisionally excited emission from Oxygen and Nitrogen~\citep{Ost89}
and recombining Hydrogen~\citep{Hum94}, and neutral gas cooling is
assumed to follow Newton's Law of exponential cooling with a cooling
time $t_c=10\,$kyr.  Photoionisation assumes a monochromatic ionising
source with $5.0\,$eV of thermal energy added to the gas per
ionisation.  Collisional ionisation is calculated using the rate
of~\citet{Vor97}.

\subsection{H~\textsc{II} region expansion with a magnetic field}
\label{ssec:KSG07}
\citet[][hereafter KSG07]{KruStoGar07} added a photoionisation module
to the \textsc{Athena} code to calculate the 3D R-MHD expansion of an
H~\textsc{II} region into a uniform medium with an initially uniform magnetic
field.  We have modelled this problem both in axisymmetry and in 3D to
compare results from our code to those of KSG07. Our 3D model used a
smaller domain of diameter $14\,$pc resolved by $160^3$ cells (12 per
cent larger cell diameters than KSG07). The only other difference in
our simulation was to add a low level of noise to the otherwise
uniform initial density and gas pressure fields.  An initial ambient
temperature of $T=100\,$K was used rather than the $11\,$K used by
KSG07 and we used the C2 cooling function described above, but tests
showed this does not affect the results significantly.

\begin{figure}%[ht]
  \centering 
\includegraphics[width=0.48\textwidth]{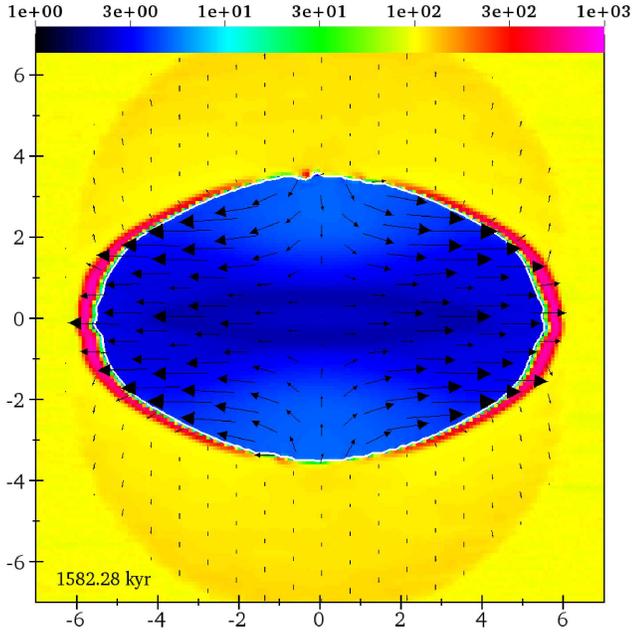}
\caption{R-MHD model of 3D H~\textsc{II} region expansion.  Results
  are shown at $1.58\,$Myr for a slice through the symmetry axis of
  the 3D simulation. Coordinate axes show position in parsecs, and gas
  number density is shown on a logarithmic scale from
  $n_{\mathrm{H}}=1$--$10^3\,\mathrm{cm}^{-3}$ as indicated.  Velocity
  vectors are also shown where vector size is linearly proportional to
  velocity, $v$.  The maximum velocity is
  $v_{\mathrm{max}}=6.7\,\mathrm{km}\,\mathrm{s}^{-1}$ in this figure.
  A white contour marking ion fraction $x=0.5$ is also shown.  The
  $14.2\,\mu$G magnetic field is initially horizontal (parallel to the
  symmetry axis).}
  \label{fig:KSG07}
\end{figure}

Gas density and velocity field are shown for a slice through the
$x$--$y$ plane at $t\simeq1.58\,$Myr in Fig.~\ref{fig:KSG07}.
Comparison to fig.~18 of KSG07 shows that very similar results are
found here, notably the dense shell which forms for expansion in the
$\pm\hat{\mathbf{x}}$ directions and its absence in perpendicular
directions.  We also find the instabilities which were noted by KSG07
in the perpendicular direction; we suggest they are real (i.e.~not
numerical artefacts) and related to the R-HD ionisation front
instabilities studied by previous authors~\citep[e.g.][]{GarSegFra96,
  Wil02, WhaNor08} on the basis of tests which showed the
instabilities are only present when neutral gas can cool strongly.
The velocity arrows show that the magnetic field significantly alters
the ionised gas dynamics: at late times the strong outflow along field
lines transforms the ionisation front to a recombination front in the
$\pm\hat{\mathbf{x}}$ directions.  We plan to study these features in more
detail in future work.

% ***************************************************
\section{3D R-MHD simulations}
\label{sec:simulations}
% ***************************************************
To focus specifically on the effects of magnetic fields on the
photoionisation process, the dense clump configurations in models 17
and 18 from ML10 were used rather than the random clump distributions
in models 1--16.  The two clump configurations as well as the
simulation domain and radiation source properties are described in
Table~\ref{tab:clump_props}.  Configuration 1 is basically the same as
model 17 in ML10; configuration 2 is changed slightly -- the two
front clumps are slightly larger and further apart, and an extra clump
is added further from the source.  The radiation source located at the
origin has the same properties as before: an ionising photon
luminosity $L_{\gamma}=2\times10^{50}\,\mathrm{s}^{-1}$ with monochromatic
photon energy $h\nu_0-13.6\,\mathrm{eV}=5.0\,$eV.  The simulation
domain used here is twice as large in every dimension as was used for
models 17 and 18 in ML10; this is because boundary effects were found
to be much more significant with strong magnetic
fields \citep[cf.][]{HenArtDeCEA09}.  The simulation domain contains
$384\times256^2$ cells, giving a physical resolution of $\Delta x
\simeq 0.012\,$pc per cell.  Results of 3D R-MHD simulations with
various domain sizes are described in Appendix~\ref{app:DomainSize}
and demonstrate that the domain used here is sufficient to prevent
significant boundary effects for all but the strongest fields modelled.

\begin{table}
  \centering
  \begin{tabular}{| l | l | c  c  c  c  c  c |}
    \hline
    Config. & Object & $x$ & $y$ & $z$ & $\delta_0$ & $r_0$ & M\\ 
    \hline
    1,2 & $X_{\mathrm{min}}$ & 1.5 & -1.5 & -1.5 &  n/a & n/a & n/a \\
    1,2 & $X_{\mathrm{max}}$ & 6.0 &  1.5 &  1.5 &  n/a & n/a & n/a \\
    1,2 & Source &0&0&0&  n/a & n/a & n/a \\
    \hline
    1 & Cl 1 & 2.30 & 0    &   0     & 500 & 0.09 & 28.4 \\
    1 & Cl 2 & 2.75 & 0    &   0.12  & 500 & 0.09 & 28.4 \\
    1 & Cl 3 & 3.20 & 0    &  -0.12  & 500 & 0.09 & 28.4 \\
    \hline
    2 & Cl 1 & 2.30 & 0    & -0.174 & 250 & 0.12 & 33.7 \\
    2 & Cl 2 & 2.30 & 0    &  0.174 & 250 & 0.12 & 33.7 \\
    2 & Cl 3 & 2.73 & 0    &    0   & 500 & 0.09 & 28.4 \\
    2 & Cl 4 & 4.20 & 0.12 &  0.12  & 400 & 0.12 & 53.9 \\
    \hline
  \end{tabular}
  \caption{Source and clump properties in R-MHD simulations.
    Configuration 1 is model 17 from ML10 and configuration 2 is a
    modified version of model 18 with an extra clump added further
    from the source.  The domain minimum/maximum coordinates are
    measured in parsecs relative to the source, which is at the
    centre of the $y$--$z$ domain.  Clump overdensities, $\delta_0$,
    are relative to the background number density of
    $n_{\mathrm{H}}=200\,\mathrm{cm}^{-3}$; their Gaussian scale
    radii, $r_0$, are in parsecs and their total masses, $M$, in units
    of \msun.}
  \label{tab:clump_props}
\end{table}

These models were run with zero-gradient boundary conditions, which
perfectly model supersonic outflow and damp reflected waves in
subsonic outflow. There is nothing to stop inflow, however, if that is
what the solution tends towards, and some of the R-MHD simulations do
develop strong inflows at later times. An alternative boundary
condition was imposed in the 2D slab-symmetric R-MHD simulations
of~\citet{Wil07}: when the velocity was outward the usual
zero-gradient boundary was applied, but if the velocity changed sign
then inflow was suppressed. This `only-outflow' boundary condition
was also tested in our simulations to study how significantly inflows
affected the solution, and in particular its observable
properties. This was initially tested using axisymmetric simulations
with a magnetic field parallel to the radiation propagation vector;
the results showed that for a weak or medium field up to
$B\sim50\,\mu$G the inflow made little difference. Its only effect was
to confine the photoevaporation flow from the clump to a smaller
volume by the inflow's ram pressure.  The higher mean density of
ionised gas (and associated stronger recombination) had no discernible
dynamical effect.  With a strong field of $B\simeq160\,\mu$G, however,
the inflow could overrun the photoevaporation flow and the ionisation
front, transforming it into a recombination front and hence strongly
affect the solution. For this reason the strong field simulation R8
described below was run with both a zero-gradient (as R8) and an
only-outflow (as R8a) boundary condition in the direction back towards
the radiation source.

\subsection{Description of simulations}
The simulations are listed in Table~\ref{tab:rmhd_Fields}.  R1 and R11
are R-HD simulations of clump configurations 1 and 2 respectively,
while the other simulations are R-MHD calculations of the same clump
configurations with varying field strengths and orientations.  The
field strengths were approximately $B\simeq18,\ 53,$ and $160\,\mu$G
and are referred to as weak, medium, and strong respectively.  R2--R10
are R-MHD simulations of configuration 1: R2--R4 are weak field models,
R5--R7 have a medium field, and R8--R10 a strong field.  R12--R15 are
R-MHD simulations of configuration 2: R11--R14 have a weak field and
R15 a medium field.

For the field orientation, `parallel' denotes parallel to the
radiation propagation vector along the centre of the simulation
($\mathbf{B}=B_0\hat{\mathbf{x}}$), and `perpendicular' denotes a field in
the $y$--$z$ plane.  Field orientations were chosen to be either
parallel or perpendicular, or oriented $80^{\circ}$ from the $x$-axis and
$50^{\circ}$ from the $y$-axis in the $y$--$z$ plane.  The simulations with
the latter field configuration (R3, R4, R7, R10) produced very similar
results to the perpendicular models (R2, R5, and R8) for configuration
1 so they were not run for as long as the other models and were not
run at all for configuration 2.

Whether a field is weak or strong is largely determined by its
dynamical importance, set by the plasma parameter $\beta\equiv 8\pi
p_g/B^2$, the ratio of thermal to magnetic pressure.  Since ionised
gas in H~\textsc{II} regions is largely isothermal, as is dense molecular gas,
the thermal pressure is approximately proportional to density.  Hence
a field of $50\,\mu$G could be strong or weak depending on whether the
gas is ionised or neutral, and on its density.  The plasma parameter
is shown in Table~\ref{tab:PlasmaBeta} for a range of gas pressures
encountered in the photoionisation simulations: (1) the initial
conditions have a constant pressure of
$p_g=1.38\times10^{-11}\,\mathrm{dyne}\,\mathrm{cm}^{-2}$ (or
$p_g/k_{\mathrm{B}}=10^5\,\mathrm{cm}^{-3}\,\mathrm{K}$); (2) gas at the
background density of $n_{\mathrm{H}}=200\,\mathrm{cm}^{-3}$ and the ionised gas
temperature of $T\simeq8\,000\,$K has
$p_g=4.42\times10^{-10}\,\mathrm{dyne}\,\mathrm{cm}^{-2}$, and (3) the peak
pressure typically encountered in the simulations is
$p_g\sim10^{-8}\,\mathrm{dyne}\,\mathrm{cm}^{-2}$.  For weak field simulations
the gas pressure clearly dominates and for the strong field the
situation is reversed; for the medium field case the initial
conditions are magnetically dominated but, once ionised, the gas
pressure is larger.  Thus it is expected that the weak field results
will largely follow the R-HD results, the strong field simulations
should show very different behaviour, and the medium field models will
lie somewhere in between.  The gas pressures in these simulations are
comparable to those estimated for the pillars and their environment in
M16 (see the discussion in ML10), but it should be borne in mind that
other massive star-forming regions can have significantly higher (or
lower) gas pressures.

\begin{table}
  \centering
  \begin{tabular}{ | l | l  l | c  c  c | c |}
    \hline
    Model & Cl. & BCs & $B_x$ & $B_y$ & $B_z$ & $t_{\mathrm{sim}}$ \\
    \hline
    R1  & 1 & 1 & 0 & 0 & 0 & 800 \\
    R2  & 1 & 2 & 1.77 & 1.77 & 17.7 & 525 \\
    R3  & 1 & 1 & 3.1 & 13.4 & 11.2 & 300 \\
    R4  & 1 & 2 & 3.1 & 13.4 & 11.2 & 300 \\
    R5  & 1 & 1 & 0 & 0 & 53.2 & 800 \\
    R6  & 1 & 1 & 53.2 & 0 & 0 & 800 \\
    R7  & 1 & 1 & 9.23 & 40.1 & 33.7 & 450 \\
    R8  & 1 & 1 & 14.2 & 7.1 & 158.9 & 400 \\
    R8A & 1 & 3 & 14.2 & 7.1 & 158.9 & 400 \\
    R9  & 1 & 1 & 159.5 & 7.1 & 10.6 & 800 \\
    R10 & 1 & 1 & 27.7 & 120.3 & 101.0 & 280 \\
    \hline
    R11 & 2 & 3 & 0 & 0 & 0 & 700 \\
    R12 & 2 & 3 & 1.77 & 1.77 & 17.7 & 375 \\
    R13 & 2 & 3 & 1.77 & 17.7 & 1.77 & 700 \\
    R14 & 2 & 3 & 17.7 & 1.77 & 1.77 & 700 \\
    R15 & 2 & 3 & 5.3 & 5.3 & 53.2 & 525 \\
    \hline
  \end{tabular}
  \caption{Clump and initial magnetic field configurations, boundary conditions
    applied, and final simulation times for the R-MHD simulations.  The
    two clump configurations indicated in column 2 are described in
    Table~\ref{tab:clump_props}.  For the boundary conditions (BCs,
    column 3), 1 refers to zero-gradient, 2 refers to only-outflow,
    and 3 is a hybrid where all boundaries are zero-gradient except
    for the boundary facing the 
    source which is only-outflow.  Field strengths are given in units of
    $\mu$Gauss.  Some models have slightly
    off-axis field orientations to prevent any potential
    grid-alignment numerical effects.  The simulations were evolved to
    the time shown by $t_{\mathrm{sim}}$ (in kyr).}
    \label{tab:rmhd_Fields} 
\end{table}

\begin{table}
  \centering
  \begin{tabular}{| l | l  l  l |}
    \hline
    $\vert\mathbf{B}\vert$ ($\mu$G) & $p_g=1.38\times10^{-11}$, &
    $4.42\times10^{-10}$, & $10^{-8}$ \\
    \hline
    18  & 1.1   & 34   & 800 \\
    53  & 0.12  & 4.0  & 89  \\
    160 & 0.014 & 0.43 & 10 \\
    \hline
  \end{tabular}
  \caption{Plasma beta ($\beta\equiv 8\pi p_g/B^2$) values for gas pressures
    encountered in R-MHD simulations, with $\beta\ll1$ indicating
    magnetically dominated regions.  The field strengths quoted are the three
    initial field strengths applied in the simulations.  Gas pressures
    ($\mathrm{dyne}\,\mathrm{cm}^{-2}$) are representative of the initial
    conditions (2nd 
    column), ionised gas at the background density (3rd column), and
    the typical maximum pressure near the ionisation front (4th column).} 
  \label{tab:PlasmaBeta}
\end{table}

\begin{figure*}
  \centering 
  \includegraphics[width=0.7\textwidth]{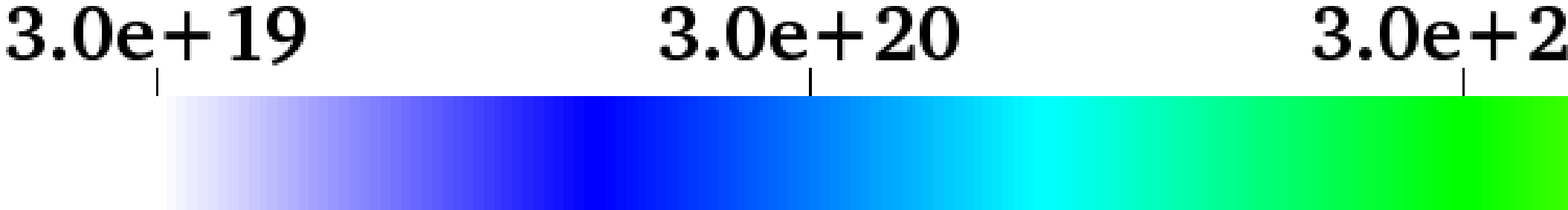}
  \includegraphics[width=0.44\textwidth]{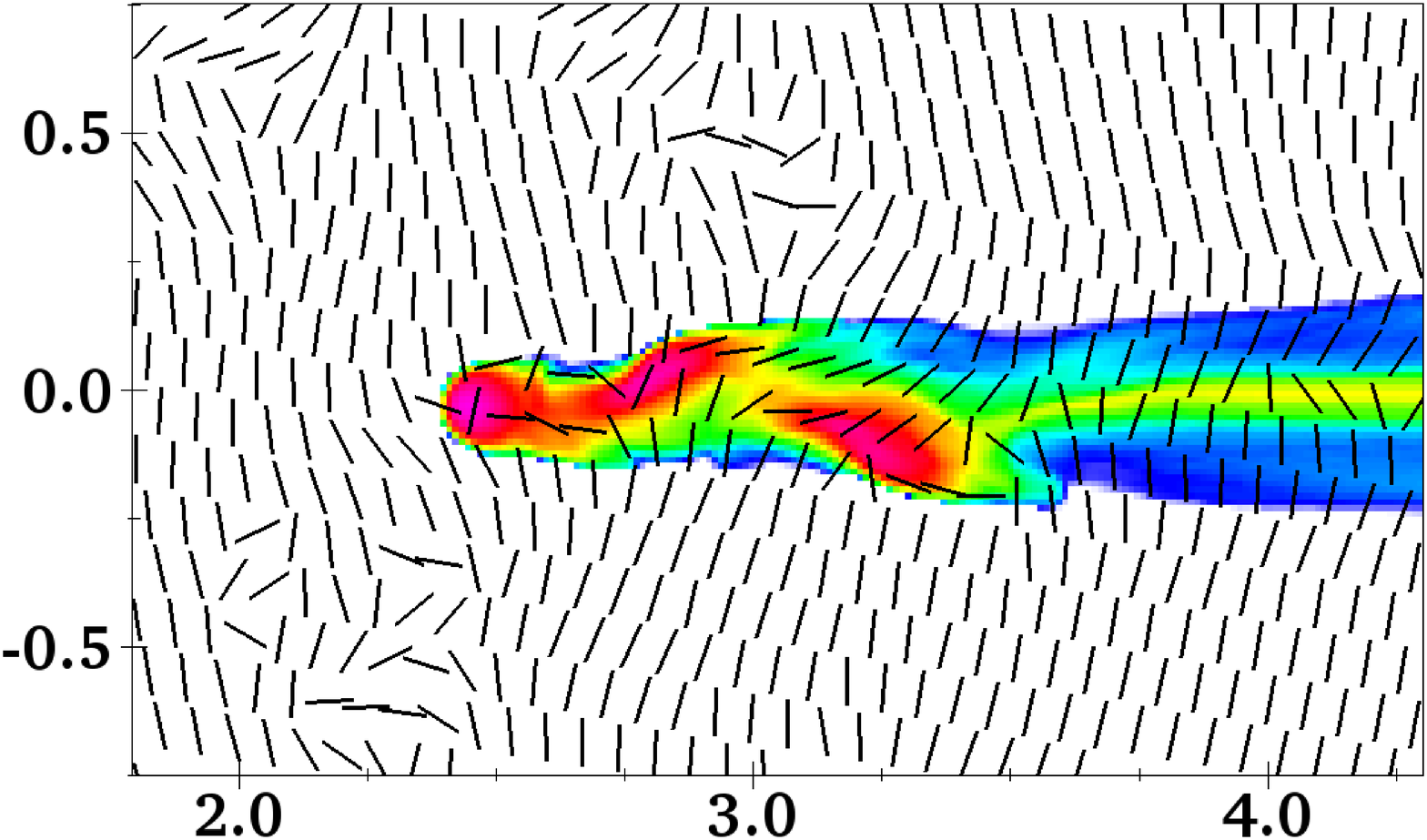}
  \includegraphics[width=0.44\textwidth]{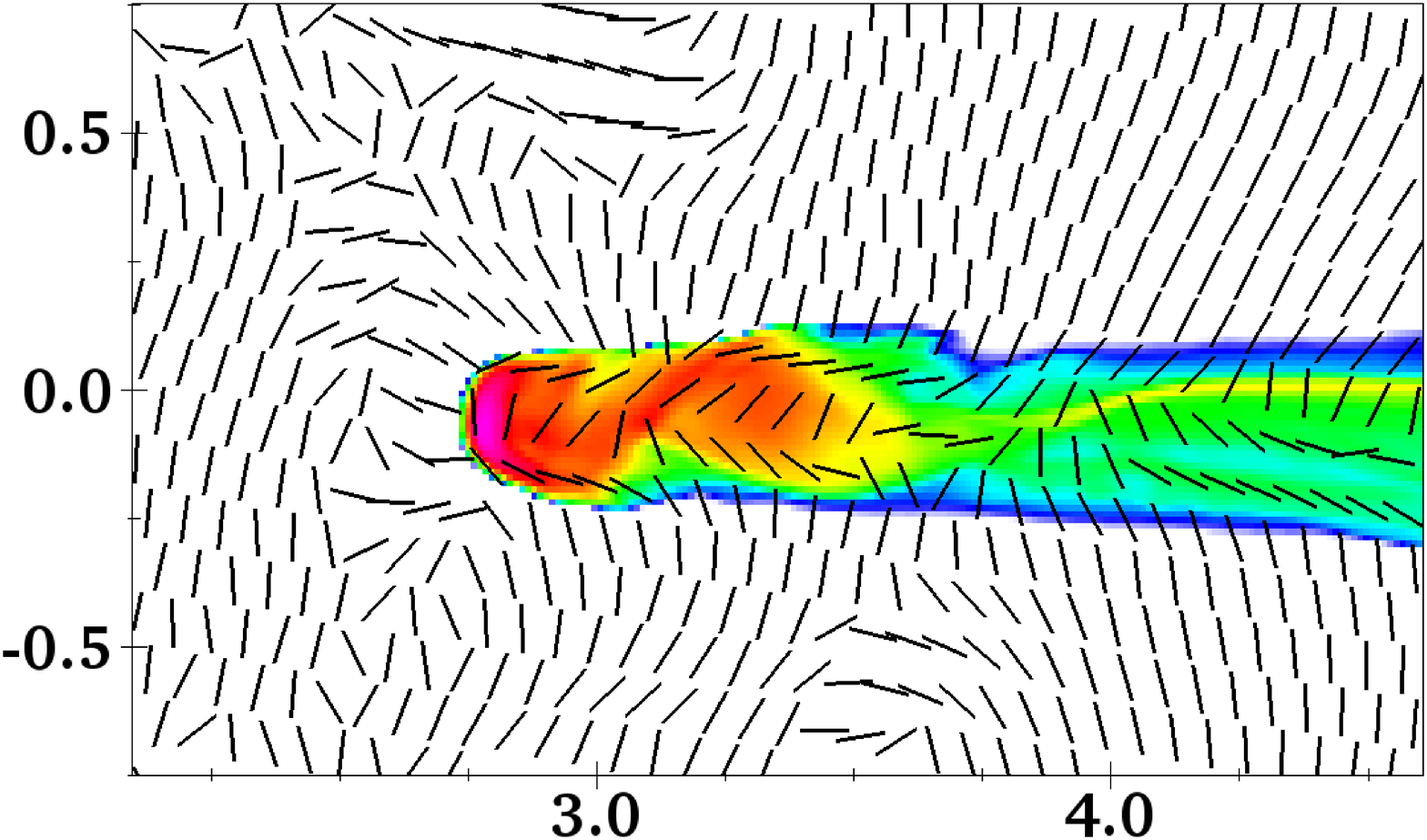}
  \includegraphics[width=0.44\textwidth]{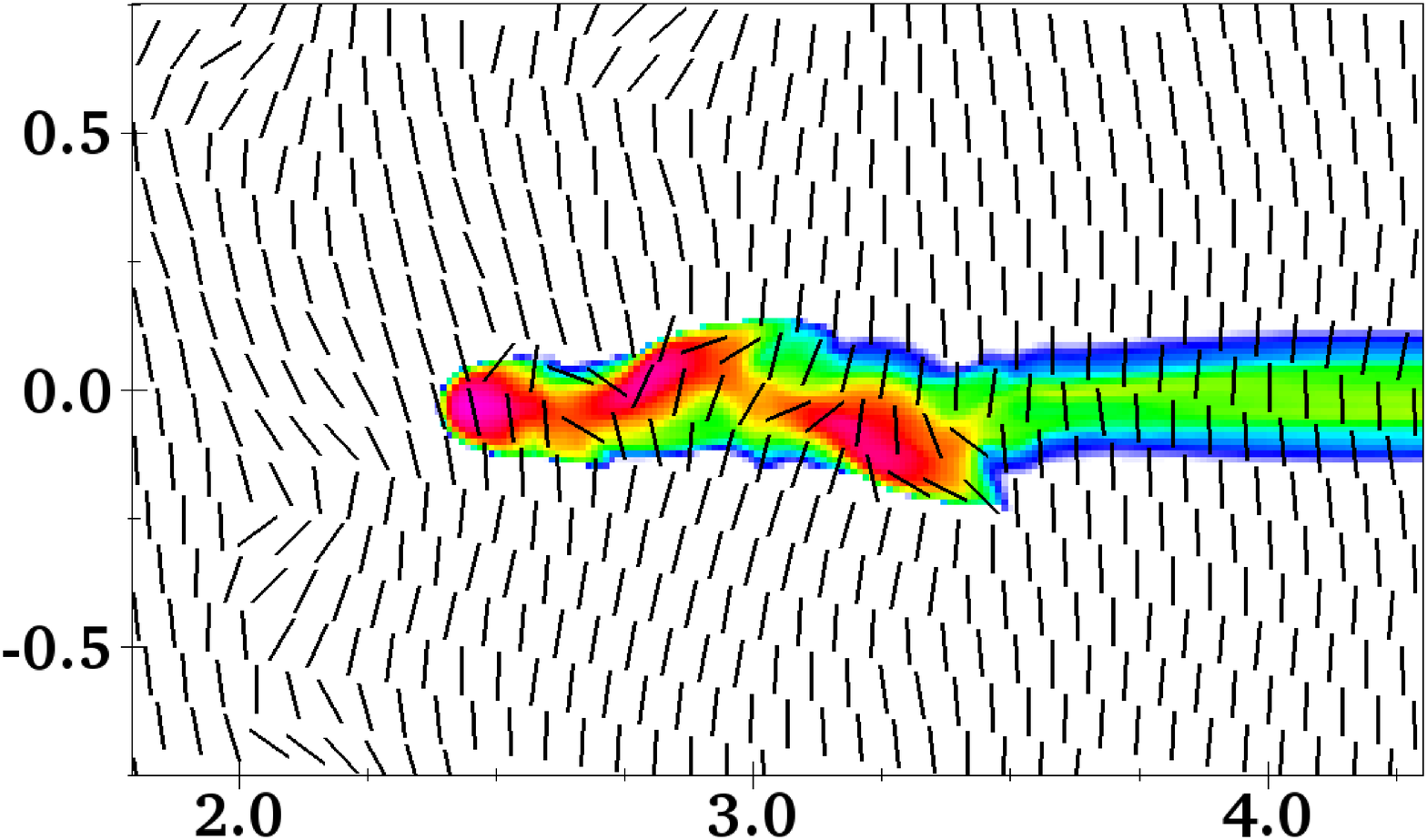}
  \includegraphics[width=0.44\textwidth]{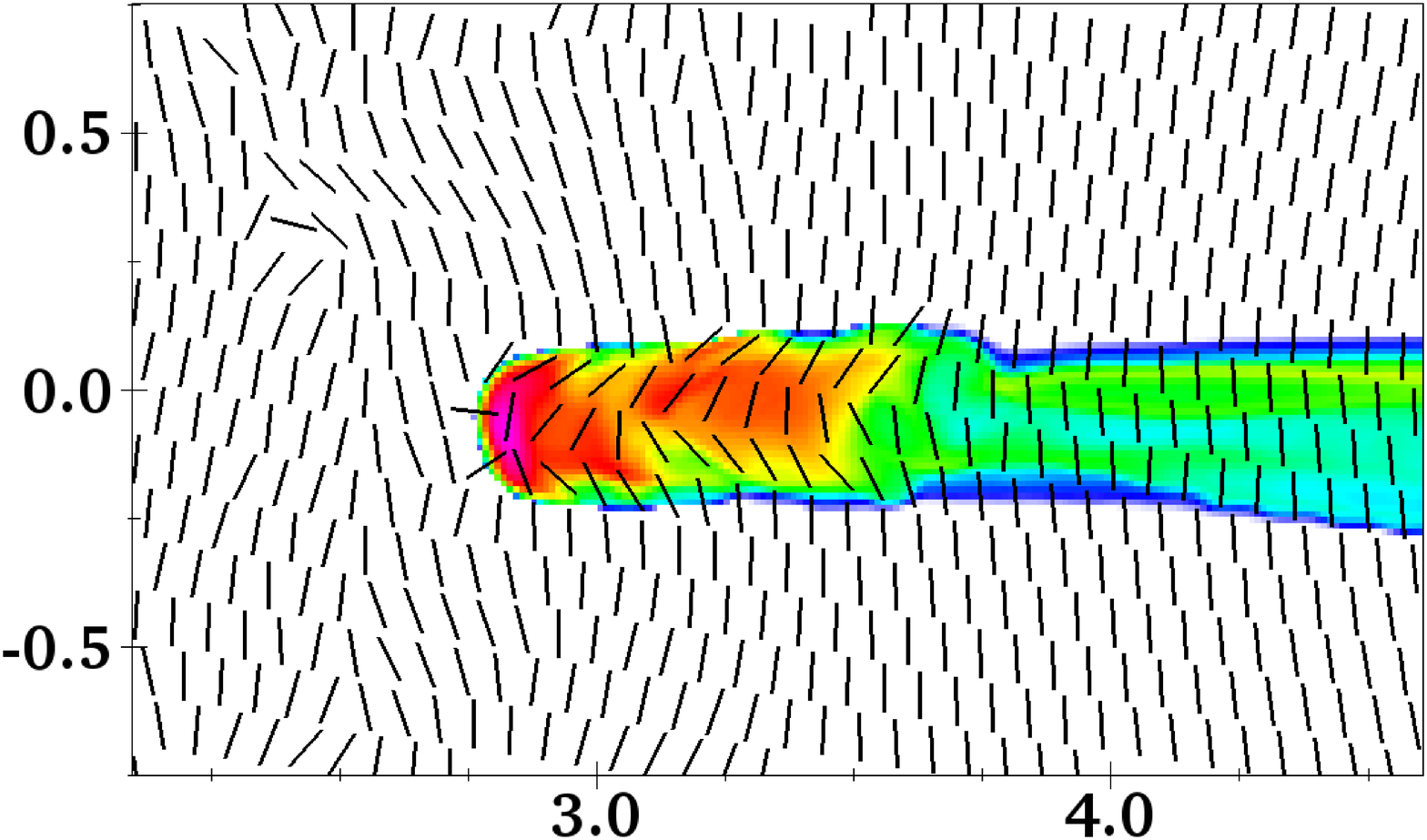}
  \includegraphics[width=0.44\textwidth]{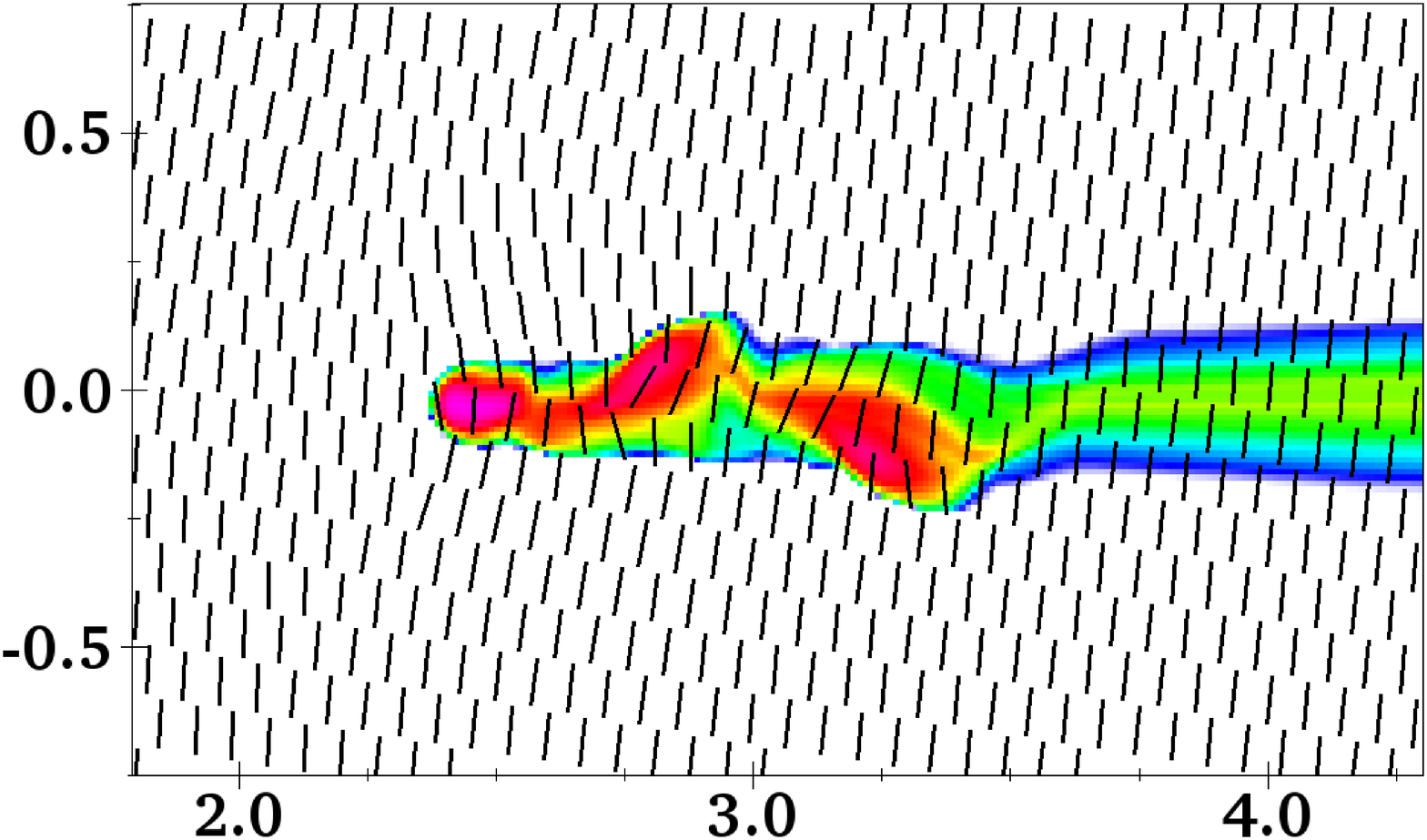}
  \includegraphics[width=0.44\textwidth]{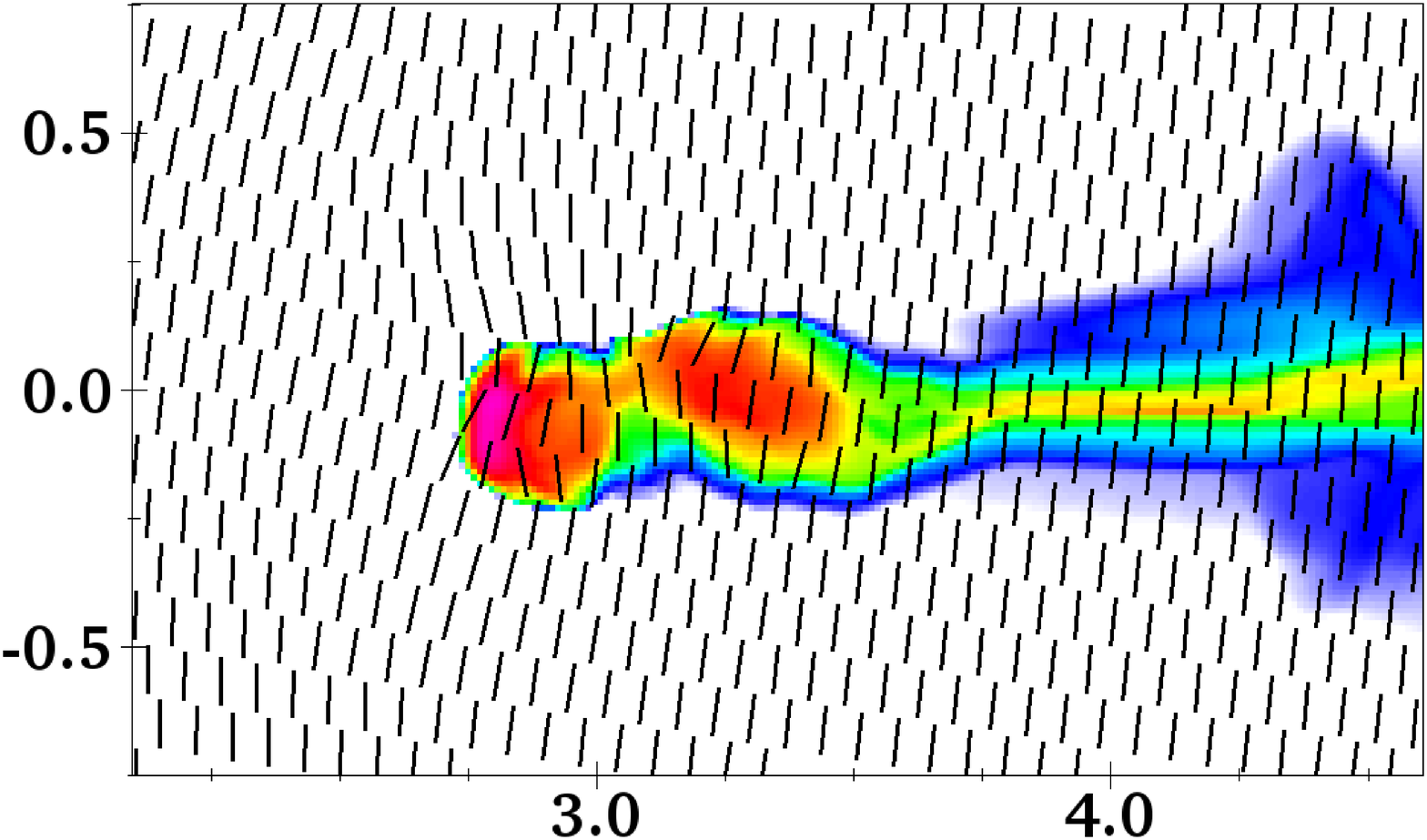}
  \caption{Projections along the $y$-axis for simulations R2 (top), R5
    (centre), and R8 (bottom) at simulation times $100\,$kyr (left)
    and $200\,$kyr (right).  Only part of the simulation domain is
    shown; positions are shown in parsecs relative to the source.
    Neutral gas column density is plotted on a logarithmic scale
    (cm$^{-2}$) as indicated.  Projected magnetic field orientation
    (Eq.~\ref{eqn:Bproj}) is indicated by the black lines.}
  \label{fig:R258_RhoB_I}
\end{figure*}

\begin{figure*}
  \centering 
  \includegraphics[width=0.7\textwidth]{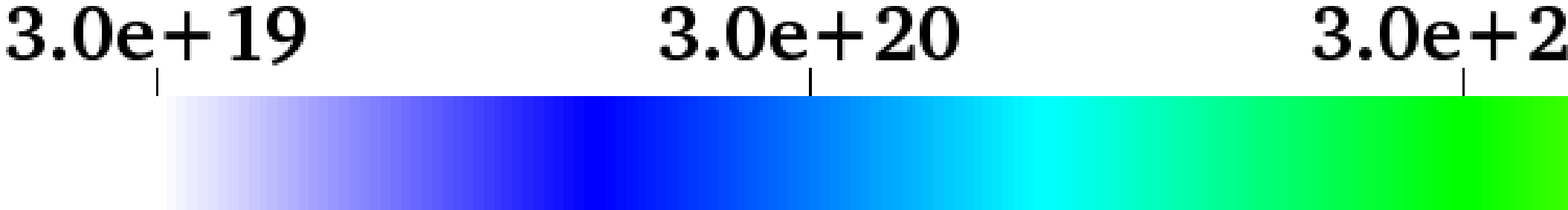}
  \includegraphics[width=0.44\textwidth]{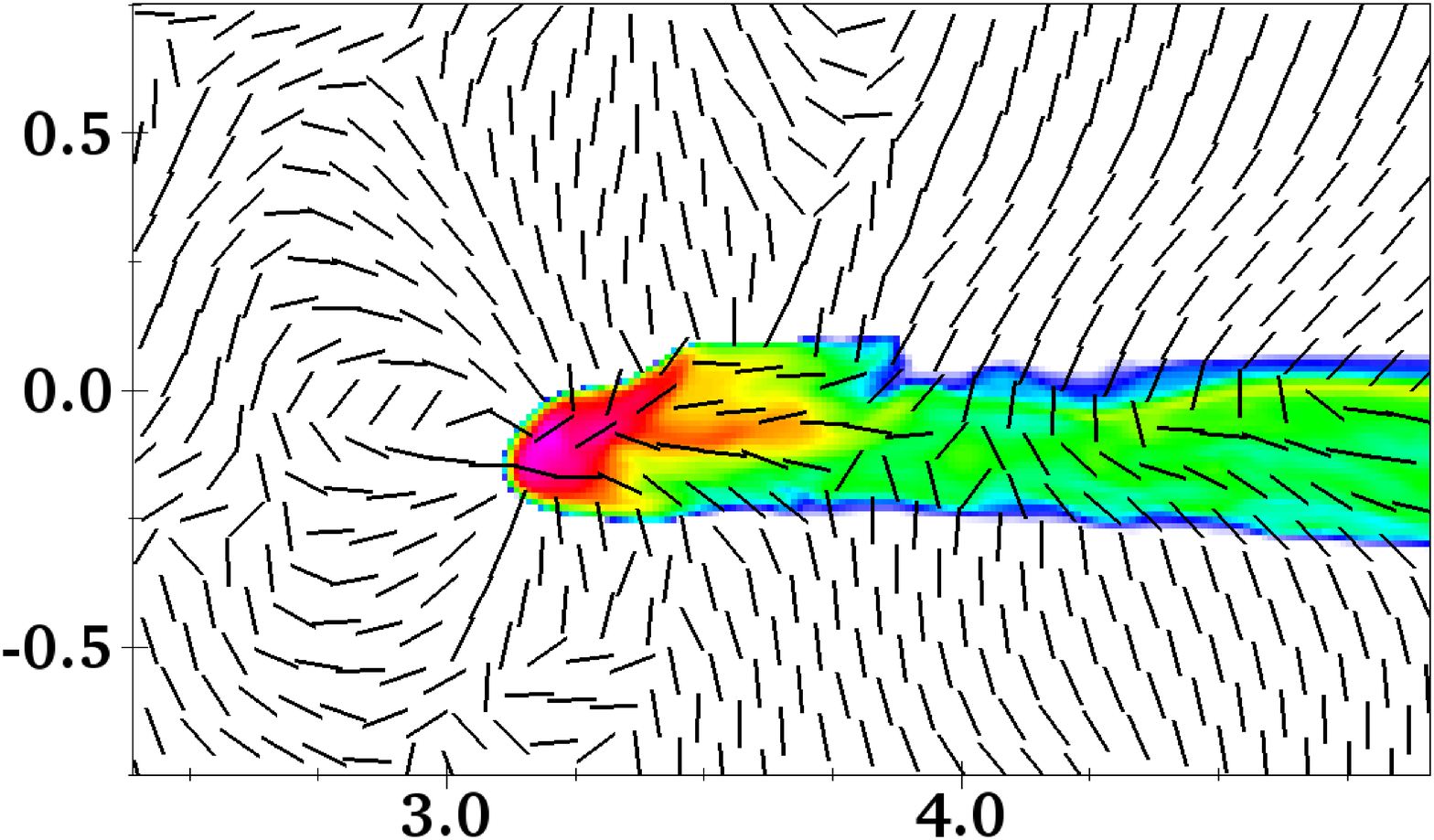}
  \includegraphics[width=0.44\textwidth]{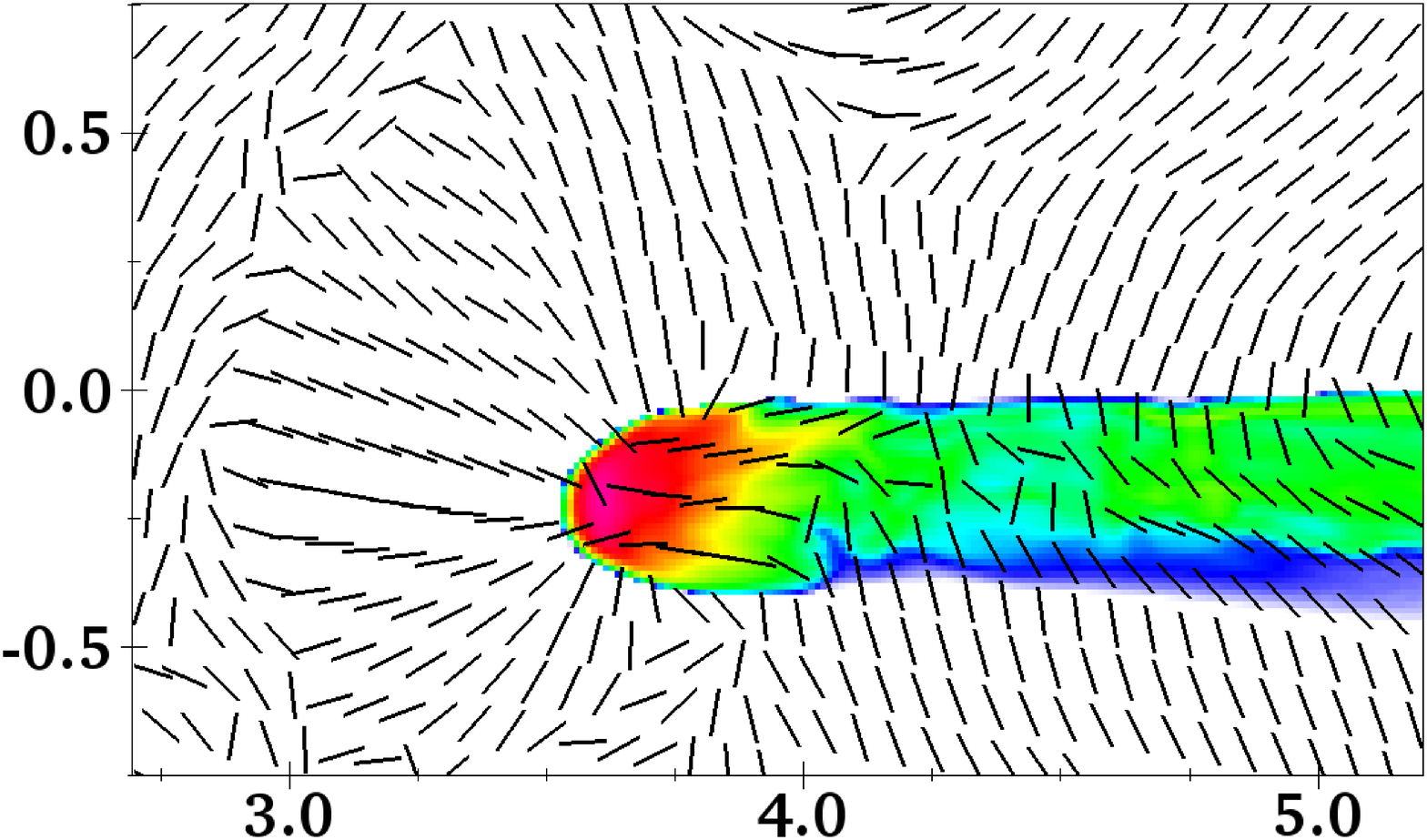}
  \includegraphics[width=0.44\textwidth]{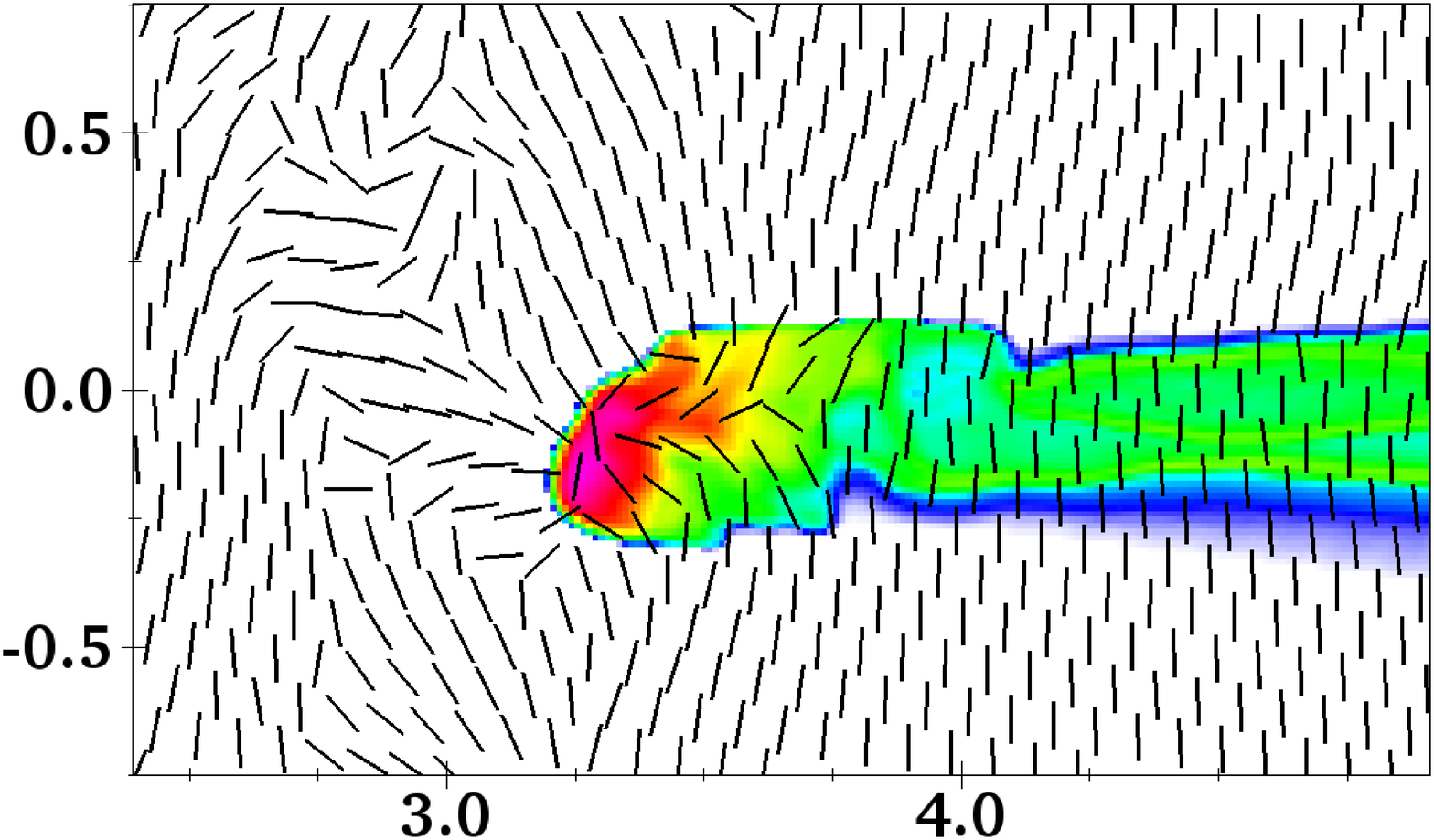}
  \includegraphics[width=0.44\textwidth]{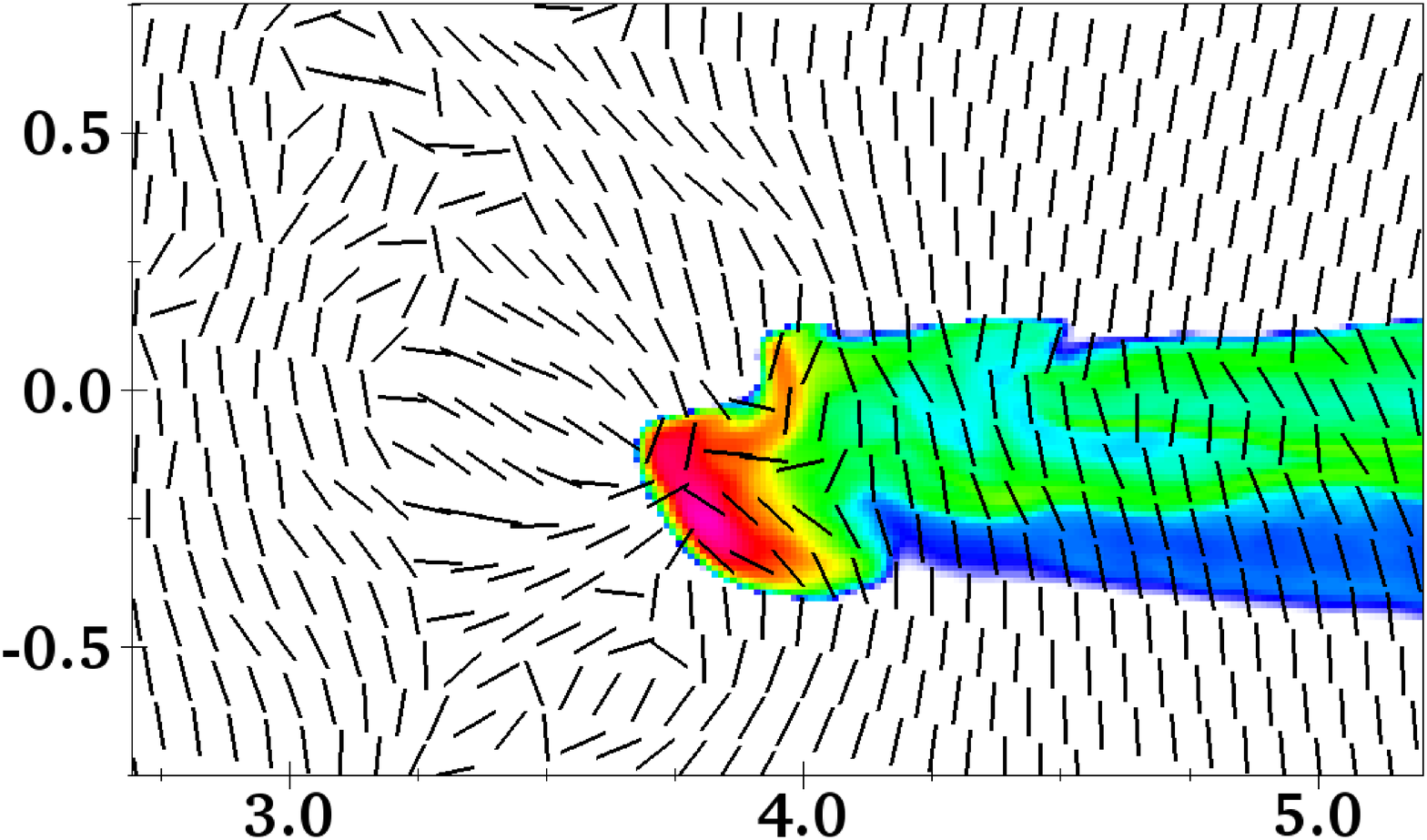}
  \includegraphics[width=0.44\textwidth]{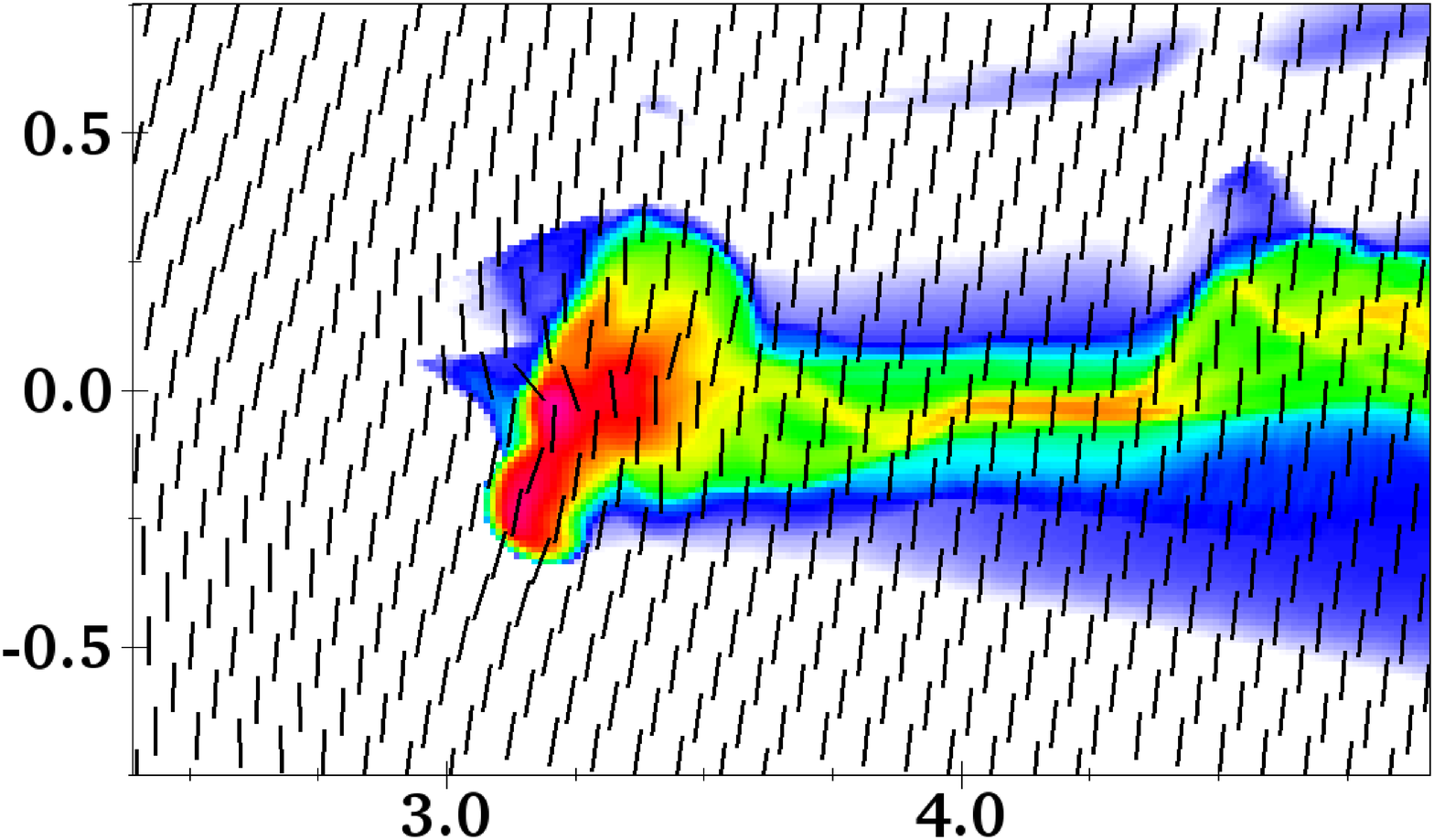}
  \includegraphics[width=0.44\textwidth]{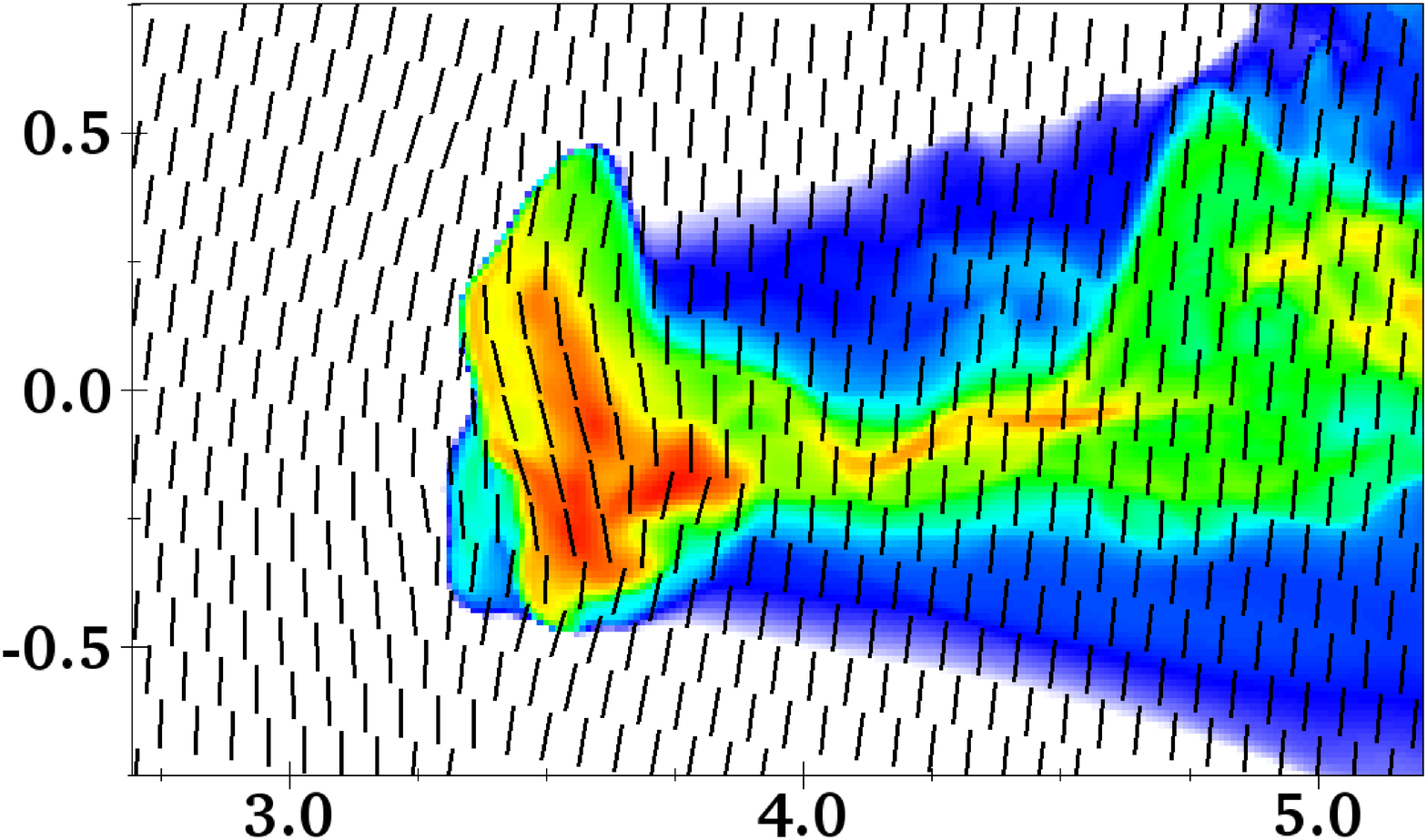}
  \caption{As Fig.~\ref{fig:R258_RhoB_I}, but for simulation times
    $300\,$kyr and $400\,$kyr.}
  \label{fig:R258_RhoB_II}
\end{figure*}

\section{Results}
\label{sec:results}
Three main observable consequences of the magnetic field are
expected \citep[cf.]{Wil07,HenArtDeCEA09}: (1) For weak magnetic
fields the field orientation will be be changed by the dynamics of the
photoionisation process. (2) When the field is sufficiently strong the
density structure of the neutral gas will be significantly altered
because gas can only move along field lines; RDI produces a sheet
rather than an axisymmetric compression. (3) Strong magnetic fields
will confine the photoevaporation flow changing its observable
properties.  We first plot projections through the simulation domain
showing column density of neutral gas and the projected magnetic
field.  Following this we show emission maps in recombination
radiation.  R2, R5, and R8 proved most useful for showing the effects
of increasing field strength so we focus most of our analysis on these
three simulations.

\subsection{Projected density and magnetic field}
\label{ssec:projections}
%\textbf{
%
The column density is calculated as in ML10 by integrating the neutral
gas number density along the line of sight (LOS).  Note that because
we do not consider molecules explicitly, the column densities should
be divided by 2 to give N(H$_2$).  The projected magnetic field is
more difficult to calculate since it must be a weighted integral, for
example to calculate the polarisation of background starlight induced
by aligned dust grains~\citep[cf.\ observations of M16
by][]{SugWatTamEA07}.  For our integration we assume a constant
gas-to-dust ratio and weight the integral by gas density.  To allow
for the possibility that grain alignment may be less effective at high
densities~\citep[e.g.][]{GooJonLadEA95} we limit the density weighting
to a maximum density,
$n_{\mathrm{max}}=2.5\times10^4\,\mathrm{cm}^{-3}$; this is rather
ad-hoc but it has a very limited effect on the projected field
orientation as long as $n_{\mathrm{max}}\gtrsim
10^4\,\mathrm{cm}^{-3}$.  The projected field may be calculated by
integrating the Stokes $Q$ and $U$ parameters along the LOS and
subsequently recovering the field orientation using trigonometric
relations~\citep[see][]{ArtHenMelEA10}.  With the approximations
described above the projected $Q$ and $U$ parameters of the magnetic
field are given by
\begin{align}
\langle Q \rangle = \int_{s=0}^{\infty}
\min[n_{\mathrm{H}}(s),n_{\mathrm{max}}]\frac{B_1^2-B_2^2}{\sqrt{B_1^2+B_2^2}}
ds \,,\nonumber\\
\langle U \rangle = \int_{s=0}^{\infty}
\min[n_{\mathrm{H}}(s),n_{\mathrm{max}}]\frac{2B_1B_2}{\sqrt{B_1^2+B_2^2}}
ds \,, \label{eqn:Bproj}
\end{align}
where $\hat{\mathbf{s}}$ is the LOS and the field perpendicular to the
LOS is $\mathbf{B}_p=[B_1,B_2]$.  Technically this integration has
units of $\mathrm{G}.\mathrm{cm}^{-2}$ but the normalisation was not
considered because we only use the projected orientation of the field
in this work and not its magnitude.  In principle the relationship
between polarisation and magnetic field is a very complex function of
temperature, density, local radiation field and dust
properties~\citep[e.g.][]{Spi98} but the microphysical processes
included in our simulations are not sufficient to model this in any
detail.
%
%}

Results from the perpendicular field models R2, R5 and R8 are shown in
Figs.~\ref{fig:R258_RhoB_I} and~\ref{fig:R258_RhoB_II} for simulation
times 100, 200, 300 and $400\,$kyr.  The LOS is the simulation
$y$-axis so the initial field in all three simulations is almost vertical and
perpendicular to the LOS.  At $100\,$kyr the leading dense
clump is just past the point of maximum compression due to RDI.  The
partially shadowed clumps are being compressed asymmetrically and are
not yet at maximum compression.  At this stage the structure of the
dense gas is very similar in all three models, although it should be
noted that the low density tail of the strong field model (R8) is
actually a thin sheet seen edge-on, whereas it is closer to
axisymmetric in R2.  Already the magnetic field has been significantly
altered in R2 due to RDI; this is less apparent in R5 while the field
in R8 is almost unaffected.  By $200\,$kyr the clumps have re-expanded
somewhat (limited because much of the compressional heating was
radiated away) but the rocket affect has taken hold and the first
clump has almost fully merged with the second.  At this stage the
magnetic field re-orientation in R2 is more pronounced and in R5 is
beginning to become significant.  There are also differences in the
morphology between the three models, quite slight between R2 and R5
but significant for R8.  After $300\,$kyr the differences have become
even more pronounced although the three clumps have basically merged
to one in all three simulations. By $400\,$kyr R2 and R5 have become
cometary globules rather than pillar-like structures.  R5 is also
fragmenting -- the small protrusion at the top of the dense region
eventually detaches from the main clump and is rapidly accelerated.
R8 at this stage looks completely different to the other models.  Low
density gas has recombined behind the very broad ionisation front
formed by gas expansion along the field lines.  The field remains
almost unchanged from its initial value even at this late stage.

\begin{figure}
\centering 
  \includegraphics[width=0.47\textwidth]{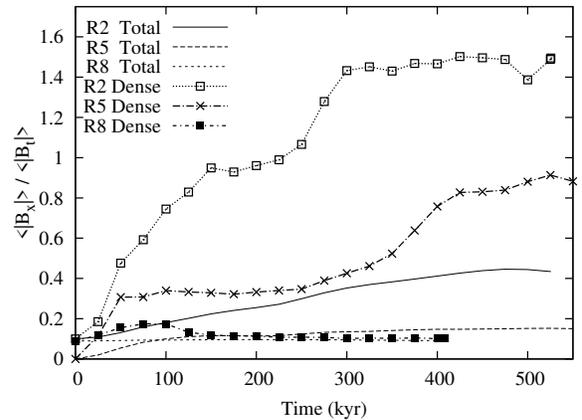}
  \caption{Evolution of the ratio of the mean parallel field
    $\langle\vert B_x\vert\rangle$ (volume averaged) to the mean
    perpendicular field $\langle\sqrt{B_y^2+B_z^2}\rangle$ for
    simulations R2, R5, R8.}
  \label{fig:R258_Bevo}
\end{figure}

The field evolution is shown in a more quantitative way in
Fig.~\ref{fig:R258_Bevo}, where the ratio of the mean (volume
averaged) parallel field $\langle\vert B_x\vert\rangle$ to
perpendicular field $\langle\sqrt{B_y^2+B_z^2}\rangle$ is plotted as a
function of time for both the full simulation domain and for only
those cells with gas density $n_{\mathrm{H}}\geq5\,000\,\mathrm{cm}^{-3}$. This ratio
changes relatively little in the full domain because ionised gas takes
up the overwhelming majority of the simulation volume. For the dense
gas, however, a very strong effect is seen in the evolution as the
field strength increases. The weak field in model R2 is swept into
alignment with the pillar by the dynamics of RDI and the rocket
effect, but a sufficiently strong field (here $\simeq160\,\mu$G in R8)
is unaffected.

In the parallel field models R6 and R9 there was very little field
evolution because the rocket effect acts in the same direction as the
initial field orientation. The results obtained are comparable to the
S00L model in \citet{HenArtDeCEA09}, except that for R9 a strong
inflow from the boundary closest to the source overran the ionisation
front after $t\sim200\,$kyr, leading to a stable 1D recombination
front. This would not be expected to form if an only-outflow boundary
is imposed, as in~\citet{Wil07}, although we note that for 1D
photoionisation where ionised gas is confined to stay in the LOS back
towards the source, a recombination front is a more stable and natural
solution than an ionisation front.

\begin{figure*}
  \centering 
  \includegraphics[width=0.8\textwidth]{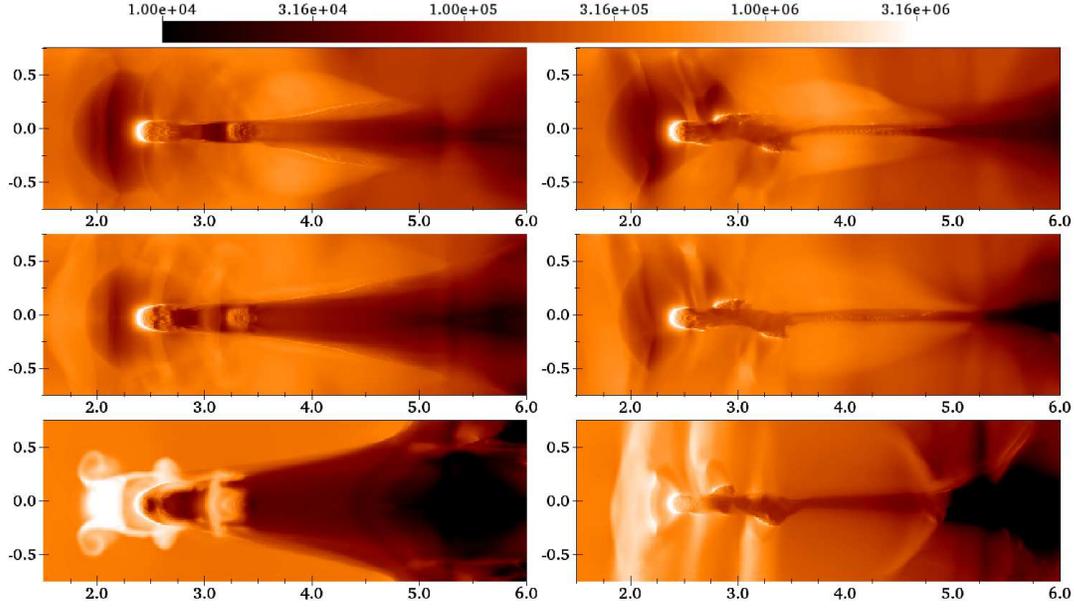}
  \caption{Projected recombination radiation maps of simulations R2
    (top), R5 (centre) and R8 (bottom) for projections along the
    $z$-axis (left) and $y$-axis (right), here at $t=100\,$kyr.  The
    initial field is along the $z$-axis for all three simulations.
    Energy flux on the indicated logarithmic scale has an arbitrary
    normalisation.}
  \label{fig:Emission100}
\end{figure*}

\begin{figure*}
  \centering 
  \includegraphics[width=0.8\textwidth]{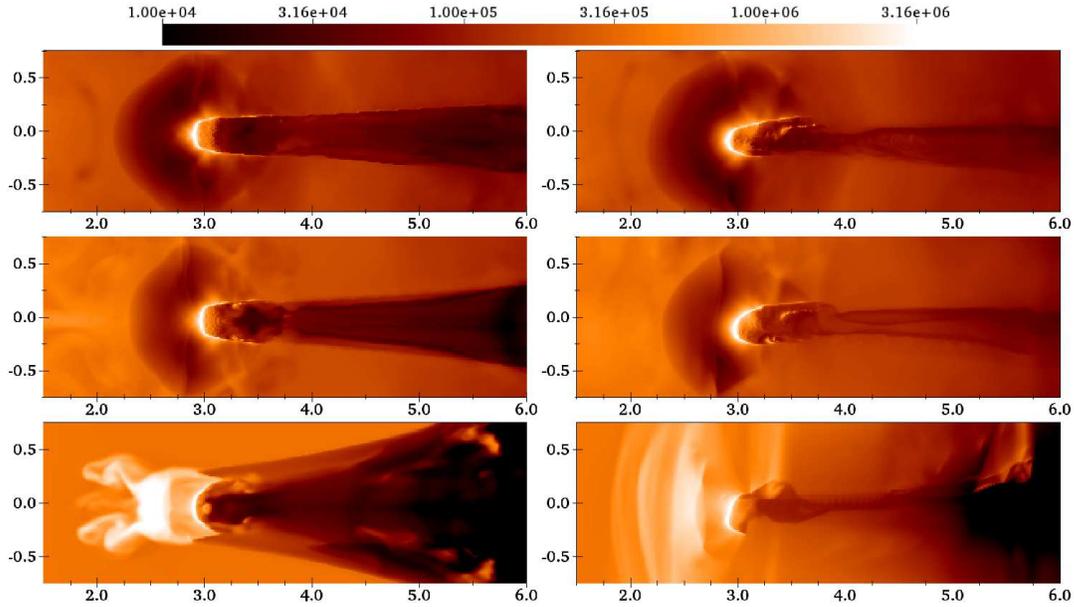}
  \caption{As Fig.~\ref{fig:Emission100} but at $t=250\,$kyr.}
  \label{fig:Emission250}
\end{figure*}

\begin{figure*}
  \centering 
  \includegraphics[width=0.8\textwidth]{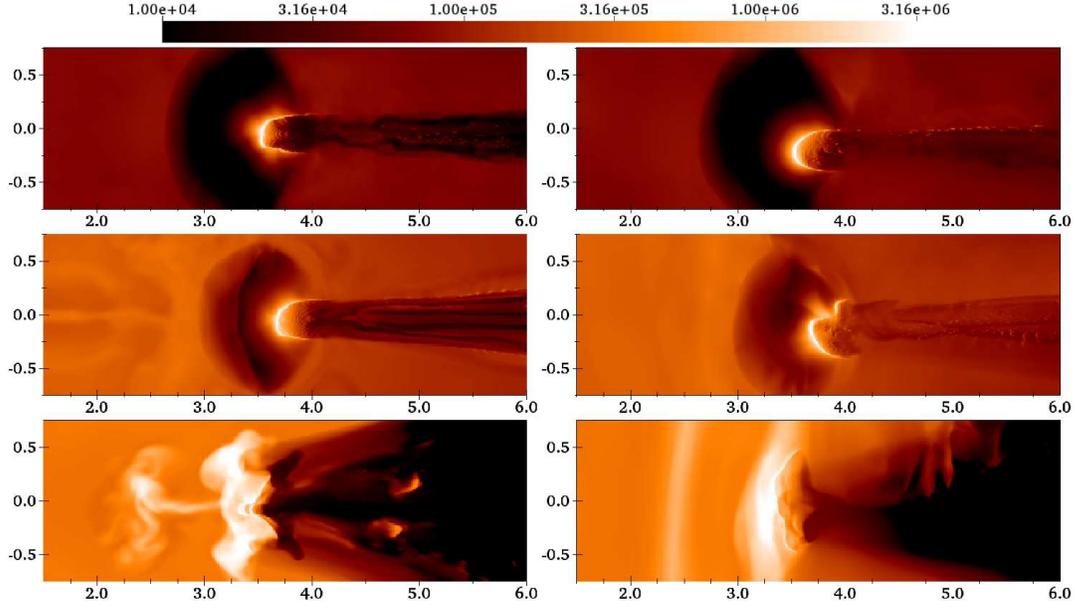}
  \caption{As Fig.~\ref{fig:Emission100} but at $t=400\,$kyr.}
  \label{fig:Emission400}
\end{figure*}

\begin{figure*}
  \centering 
  \includegraphics[width=0.8\textwidth]{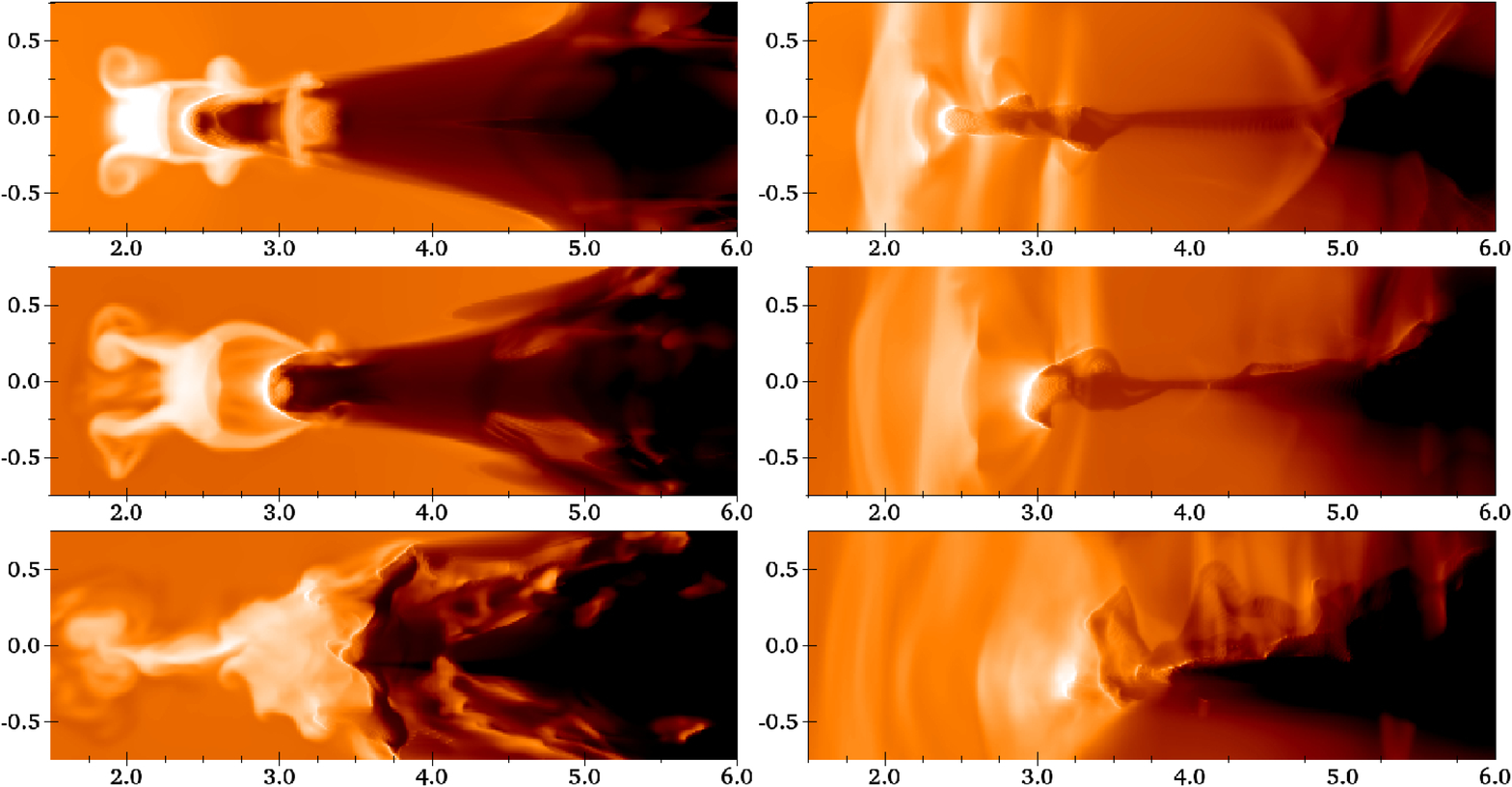}
  \caption{Similar to Fig.~\ref{fig:Emission100}, but showing
    simulation R8a at $t=100\,$kyr (top), $t=250\,$kyr (middle) and
    $t=400\,$kyr (bottom).  The left-hand plots show projection along
    the $z$-axis and the right-hand plots along the $y$-axis.  These
    can be compared to R2, R5, and R8 in
    Figs.\ \ref{fig:Emission100},\ \ref{fig:Emission250}
    and\ \ref{fig:Emission400}.}
  \label{fig:EmissionR88a_I}
\end{figure*}

\subsection{Emission maps in recombination lines}
\label{ssec:emission}
The gas emission in recombination radiation (e.g.~$\mathrm{H}\alpha$) was also
calculated from the simulation outputs as in ML10, using the
emissivity formula from~\citet{HenArtGar05} and dust opacity as
in~\citet{MelArtHenEA06} to attenuate radiation along a LOS.  Emission
maps are shown at times 100, 250, and $400\,$kyr for models R2, R5,
and R8 in Figs.~\ref{fig:Emission100},~\ref{fig:Emission250}
and~\ref{fig:Emission400}.  Here, as in the column density maps, there
are small differences between R2 and R5, but R8 shows very different
emission.  This can be understood from Table~\ref{tab:PlasmaBeta}
which shows that only R8 is magnetically dominated in ionised gas.
The initial field is along the LOS for the left-hand projections (LOS
is simulation $z$-axis) and vertical for the right-hand projections
(LOS is simulation $y$-axis).

The field is so weak in R2 that the evolution is very similar to the
purely hydrodynamical evolution in R1 (cf.~ML10) except that the low
density shadowed tail region is not quite axisymmetric.  A strong
photoevaporation flow from the front clump expands spherically with
velocity $v\simeq30$--$35\,\mathrm{km}\,\mathrm{s}^{-1}$ until the ram
pressure equals the total pressure of the ambient medium at which
point a standing shock is established.  For the full evolution of this
model the dense gas at the ionisation front is by far the brightest
structure in recombination emission.  There are small differences in
model R5, notably that the photoevaporation flow is confined to a
smaller volume by the higher ambient total pressure, but the evolution
is largely the same.

In R8 the total pressure of the ambient medium is magnetically
dominated and much higher than in R2 or R5.  This confines the
photoevaporation flow much more effectively so that the region of
spherical expansion is significantly smaller.  Instead shocked photoionised 
gas is diverted along field lines into a bar-shaped linear structure with a
density $2.5$--$4\times$ the background density. 
In \citet{HenArtDeCEA09}
this structure is termed a \emph{dense ribbon} to avoid confusion
with photon-dominated regions such as the Orion Bar \citep[see
e.g.][]{ODel01}.  We will avoid the term `bar' for the same reason,
instead referring to the structure as a ribbon or ridge
\citep[cf.][]{BohTapRotEA04}.
This dense ionised gas is quite spatially extended and so has a comparable
recombination radiation intensity to the bright rim at the ionisation
front.  Because of the high column density in the ribbon there is
significant recombination in its shadow at later times.  This
evolution is similar to that seen in the strong field models of
\citet{HenArtDeCEA09}.

The left-hand plots for R8 (with $\mathbf{B}$ along the LOS)
show wave-like features which appear to be partially developed
Kelvin--Helmholz instabilities.  These move back towards the
midplane ($y=0$) due to magnetic tension and have collided by
$t=400\,$kyr (Fig.~\ref{fig:Emission400}); similar but more
pronounced evolution was found by \citet{HenArtDeCEA09} in their simulations.

In simulation R8 at 400 kyr the dense ridge/ribbon has become
optically thick, enhancing recombination in its shadow and
dramatically weakening the photoevaporation flow.  The result of
this is that the standoff distance between the ridge and the
original ionisation front of the dense clump has decreased almost to
zero.  This was found to be a short-lived phase in the models of
\citet{HenArtDeCEA09} and we expect the same to be true here because
a strong photoevaporation flow is required to generate the overdense
ionised gas. 

\subsection{Boundary effects for simulation R8}
\label{ssec:R8R8a}
It was found that a reasonably strong inflow of gas developed at the
boundary $x=1.5\,$pc in R8 at late times, ranging from
$v_x=1$--$3.5\,\mathrm{km}\,\mathrm{s}^{-1}$. The ram pressure from this
inflow could raise the density in the shocked region and hence affect the
resulting evolution, so the simulation was repeated with this boundary
set to only-outflow, thereby preventing the inflow developing. This
simulation, denoted R8a, is shown in Fig.~\ref{fig:EmissionR88a_I} at
times 100, 250 and $400\,$kyr, again in recombination line emission.
Comparison to Figs.\ \ref{fig:Emission100} and\ \ref{fig:Emission250}
shows that the evolution is almost identical up to $t=250\,$kyr except
that the bright ridge is slightly further from the ionisation front because now
there is no ram pressure confinement. At $t=400\,$kyr the differences
are more significant (see\ Fig.~\ref{fig:Emission400}): the ionised
ridge is broader and less well-defined in R8a than in R8, and while the
dense neutral gas has a similar structure there are small
differences. The general features remain the same, however, and the
confined photoevaporation flow still shields the ionisation front
significantly and outshines it in recombination radiation.

A similar but weaker inflow was also found in simulation R5. The only
effect this has on the results is that the ionised ambient medium is
denser than in R2, as seen by comparing the mean emission between the
two simulations in Figs.~\ref{fig:Emission100}--\ref{fig:Emission400}.

\section{Ionised Ridge formation}
\label{sec:BarFormation}
The presence of a bright ionised ridge/ribbon of dense gas in simulation R8 which is absent
in R1, R2, and R5 is easily understood quantitatively by evaluating
the jump conditions for the isothermal termination/reverse shock of
the photoevaporation flow (the temperature is always in the range
$7500$--$8500\,$K in the ionised gas).  If we assume the system has
reached an equilibrium, then the total pressure in the upstream (I),
shocked (II), and downstream (III) regions will be equal.  To avoid
dealing with the spherical expansion of the photoevaporation flow we
consider region I to consist only of the conditions immediately
upstream from the shock (referred to below with subscript `1').  If we
further assume that the only significant magnetic field is transverse
to the velocity of expansion we can consider only its pressure effects
and define a speed $b^2\equiv B^2/(8\pi\rho) = p_m/\rho$ so that the
total pressure in a region is
\begin{equation}
  P_{\mathrm{tot}}=p_{\mathrm{ram}}+p_g+p_m = \rho(v^2+c^2+b^2) \,
\end{equation}
being the sum of the ram, thermal and magnetic pressures respectively.

We wish to calculate the density in region II as a function only of
the upstream velocity $v_1$ and the ambient medium properties ($b_3$,
$v_3$).  Only when the ratio $\rho_2/\rho_3>1$ will a bright ribbon or
ridge be observable.  Thus the equations to be solved
are~\citep[cf.][]{deHofTel50}
\begin{equation}
k_1(v_1^2+c^2+b_1^2) = k_2 (v_2^2+c^2+b_2^2) =
(v_3^2+c^2+b_3^2) \;,
\end{equation}
where $k_1\equiv \rho_1/\rho_3$ and $k_2\equiv \rho_2/\rho_3$.  The
isothermal sound speed is $c^2\equiv p_g/\rho = kT/(\mu m_p)$, where
$\mu$ is the mean mass per particle in units of the proton mass $m_p$
and is $\mu=0.5$ in ionised gas in our models.  For $T\simeq8300\,$K
this gives $c=11.7\,\mathrm{km}\,\mathrm{s}^{-1}$.  In all simulations
the magnetic field is negligible in region I ($b_1\ll c$) and smaller
than the other terms in region II ($b_2<c$).  The flow velocity in the
reverse shock reference frame is $v_1\simeq
33$--$35\,\mathrm{km}\,\mathrm{s}^{-1}$ in all simulations giving a
Mach number $M\equiv v_1/c \simeq 3$.

The hydrodynamic jump conditions for the isothermal reverse shock are
\begin{equation}
  \frac{\rho_2}{\rho_1} = \frac{v_1^2}{c^2} = M^2\;,\qquad 
  v_2 = c/M \;.
\end{equation}
When $b_1\simeq0$ but $b_2>0$ the density jump is given by the
solution to the quadratic equation
\begin{equation}
  \left(\frac{\rho_2}{\rho_1}\right)^2 -
  \left(\frac{\rho_2}{\rho_1}\right)\frac{v_1^2+c^2}{c^2+b_2^2}
    +\frac{v_1^2}{c^2+b_2^2}  =0 \;.
\label{eqn:PressureEqm}
\end{equation}
This equation has real solutions only for some values of $b_2$, leading
to the constraint
\begin{equation}
  0 \leq \frac{b_2^2}{c^2} \leq \frac{M^2}{4}\left(1-\frac{1}{M^2}\right)^2 \;.
\label{eqn:B2limit}
\end{equation}
For $M=3$ this gives $b_2^2 \leq 16c^2/9$ but in the simulations it is
actually significantly less than this limiting value.

Using these equations it is easy to show that in the hydrodynamic
limit $k_2=M^2/(1+M^2)\simeq 0.9$ and $k_1=1/(1+M^2)\simeq 10$.  No
bright overdensity is formed, and the photoevaporation flow shocks at a
density about $10\times$ below the ambient density.  This is indeed
seen in the R-HD simulation R1 and also the weak perpendicular field
simulation R2.

With $b_1=v_3=0$ and $b_3,b_2>0$ we obtain
\begin{align}
  k_2 = \frac{b_3^2+c^2}{(c/M)^2+c^2+b_2^2} &\geq
  \frac{1+(b_3/c)^2}{\left(\frac{1}{M}\right)^2+1
    +\frac{M^2}{4}\left(1-\frac{1}{M^2}\right)^2} \,,\nonumber\\
  k_2 &\leq \frac{M^2}{1+M^2}\left(1+\frac{b_3^2}{c^2}\right)
\end{align}
where the inequalities come from the limiting values for $b_2$ given
by Eq.~\ref{eqn:B2limit}.  With $M=3$ this gives
\begin{equation}
  \frac{9}{26}\left(1+\frac{1}{\beta_3}\right) \leq k_2 
  \leq \frac{9}{10}\left(1+\frac{1}{\beta_3}\right)
\end{equation}
where $\beta_3\equiv p_g/p_m=c^2/b_3^2$.  The overdensity in the
post-shock region is therefore determined by the value of
$(1+1/\beta)$ in the ionised ambient medium, showing that the ambient
medium must be magnetically dominated to give an over-dense ridge.  For
R5 $\beta_3=4.0$ and for R8 $\beta_3=0.43$, leading to maximum
overdensities $k_2$ of 1.1 and 3.0 respectively.  Actual post-shock
densities in R5 correspond very well to this value, and in R8/R8a the
ribbon density is $n_{\mathrm{H}}\simeq 550$--$700\,\mathrm{cm}^{-3}$
depending on the time.  The ambient density is
$n_{\mathrm{H}}=200\,\mathrm{cm}^{-3}$ so the overdensity of
$k_2=2.5$--$3.5$ is again close to the predicted value.

Similarly it is easy to show that $k_1=(1+1/\beta_3)/(1+M^2)$ giving
$k_1=0.125$ and $0.33$ for R5 and R8.  According to this equation the
photoevaporation flow should shock for simulations R1, R2, R5 and R8
at $n_{\mathrm{H}}=20,\ 21,\ 25,\ 67\,\mathrm{cm}^{-3}$ respectively.  In the
simulations at $t=150\,$kyr the actual densities are
$n_{\mathrm{H}}\simeq25,\ 22,\ 31,\ 75\,\mathrm{cm}^{-3}$, whereas at $t=200\,$kyr
they are $n_{\mathrm{H}}\simeq18,\ 19,\ 30,\ 100\,\mathrm{cm}^{-3}$, in both cases
comparable to the values predicted.

This simple model has obvious limitations, in particular we have not
specified the form of the boundary between regions II and III -- it
is not a shock, but if the magnetic field strength changes
significantly then a contact discontinuity forms to equalise the total
pressure.  Also we have not considered any parallel field component
which, if present, would alter the pressure balance.  Despite this the
model quantitatively captures the properties of the over-dense ridge.
Addition of an opposing ram pressure in the ambient medium can clearly
have a similar effect to a downstream magnetic field in that it
increases the density at which the photoevaporation flow termination
shock forms (and hence also the post-shock density $\rho_2$).  The
differences between R8 and R8a
(Figs.\ \ref{fig:Emission250}--\ref{fig:EmissionR88a_I} show that
even an inflow of $v_3\simeq3\,\mathrm{km}\,\mathrm{s}^{-1}$ can have an observable effect.
Further work will be required to investigate observational differences
(possibly in the shape and velocity of gas within the ridge
and/or whether it is sheet-like or bar-like)
which could be used to distinguish between a magnetically confined
bright ridge and one which is ram-pressure confined.

% ***************************************************
\section{Discussion}
\label{sec:discussion}
% ***************************************************
This work confirms the suggestion by~\citet{SugWatTamEA07} that their
observations can in principle constrain the field strength in M16 when
combined with detailed simulations.  On the evidence of these
simulations (and those of \citealt{HenArtDeCEA09}) it appears that the
magnetic field in the region around the pillars in M16 cannot be as
large as the $160\,\mu$G field in model R8/R8a, and is likely
$\vert\mathbf{B}\vert\lesssim50\,\mu$G.  This can be deduced by
comparing observations of the field orientation in the H~\textsc{II}
region and $\mathrm{H}\alpha$ emission around the pillars with the
emission maps presented here and in \citet{HenArtDeCEA09}.  The
recombination line intensity in the images from~\citet{HesScoSanEA96}
clearly decreases with distance from the ionisation front and was
shown to be consistent with a spherically expanding photoevaporation
flow, which matches only our weak and medium field strength models.
Additionally no prominent ribbon/ridge or sheet-like features are present near
the pillars. 

Interpreting the the polarisation measurements within the pillars in
M16 as further evidence for a weak field implicitly assumes the
pillars formed via a mechanism similar to the one considered here.
Because the rocket effect always accelerates gas away from the
radiation source and into the shadowed region, we believe that this
alignment of weak fields with the pillar axis is a generic feature of
all such models, but the initial conditions for pillar formation are
not static~\citep[cf.][]{GriBurNaaEA10} and it is uncertain how well
correlated the initial magnetic field orientation would be between
dense and rarefied gas.

Compared to the simulations of~\citet{HenArtDeCEA09}, the RDI of
clumps in our models is weaker leading to less flattened
structures.  The main reason is that here the clumps are considerably
more centrally concentrated with higher initial densities, making
them less susceptible to compression.  Additionally the clump in
their model is much closer to the ionising source and subject to
higher incident radiation flux (only partially offset by the higher
source luminosity in our model), increasing the effect of RDI.
Thermal physics models and differing geometric effects due to source
proximity may also have an effect.

The weak field model W80L in \citet{HenArtDeCEA09} should be
roughly comparable to our model R5.  Fig.~\ref{fig:R258_Bevo} shows
that the ratio $\langle\vert
B_x\vert\rangle/\langle\sqrt{B_y^2+B_z^2}\rangle$ in dense neutral
gas remains at $\sim0.3$ for 300 kyr in R5, similar to the ratio in
neutral gas shown in \citet[][fig.~12]{HenArtDeCEA09} for W80L.  The
subsequent increase in this ratio in R5 may represent a later
evolutionary phase but, because of the simulation differences
already mentioned, it is difficult to quantify this.

Our results are otherwise in good agreement with those
of \citet{HenArtDeCEA09} considering the differences in the
simulation initial conditions.  We also find that that the photoionisation
of dense clumps is very different in strongly magnetised media than in
the non-magnetised case, with the dynamics becoming more planar than
axisymmetric.  Photoevaporated ionised gas tends to form ribbon-like
structures which can (transiently) have significant optical depths and lead to
recombination behind them.  Clump compression along field lines
creates flattened sheet-like structures and an ionisation front which
is far from axisymmetric.

Despite the simplistic initial conditions used here and in ML10,
qualified support for these models comes from the 3D R-HD simulations
of~\citet{GriBurNaaEA10}.  They used an isothermal self-gravitating
decaying turbulence model as the initial conditions, varying the
turbulent Mach number at which the radiation source is switched on.
In agreement with the trends we reported they also found that to form
pillars the density field was required to include both large
low-density regions and reasonably massive dense regions of sufficient
overdensity to trap the ionisation front.  Because of the dynamic
initial conditions in their simulations they were also able to show
the dependence on initial gas motions, finding that high Mach number
turbulence did not produce pillar-like structures as successfully as
models with Mach number $M\sim4$--10 because structures formed and
dispersed too rapidly.  When the Mach number was too low a sufficient
density contrast to form pillars was not obtained.
\citet{GriBurNaaEA10} also found that ionised gas pressure is the
dominant driver of the dynamics and gravity appears to play a smaller
role, an explicit assumption in our work based on previous
calculations by~\citet{EsqRag07}.

In our opinion~\citet{GriBurNaaEA10} overstate the differences between
our results and theirs -- in both models a combination of RDI and the
rocket effect reinforces pre-existing inhomogeneities in the ISM and
generates elongated structure along the radiation propagation
direction.  We do not see our results as being in conflict with those
of~\citet{GriBurNaaEA10}; rather that they have confirmed many of our
conclusions and also extended our work by studying pillar formation in
a more realistic initial density field.  These results suggest that
the initially static models considered in ML10 and in this work can
capture the essential physics of the pillar formation and evolution,
once the initial conditions of dense regions surrounded by a less
dense ambient medium have been set up.  As noted in ML10, our models
can also explain the LOS velocities seen in the pillars in
M16~\citep{Pou98,WhiNelHolEA99}, although we do not see the helical
morphology and apparent rotation found in some elephant
trunks~\citep{GahCarJohEA06}.  This may require an initially dynamic
density field since such structures seem to occur quite readily in the
simulations of~\citet{GriBurNaaEA10}.

Our simulations (and those of \citealt{HenArtDeCEA09}) suggest that
ionisation fronts with a strong perpendicular magnetic field should
have clear observational consequences in both the morphology of the
recombination emission and of the dense gas.  \citet{HenArtDeCEA09}
note that 
while the Orion Bar has a similar linear shape to the
ribbon-like structures seen here, it is associated with the main
ionisation/dissociation front itself rather than being an overdensity within the
H~\textsc{II} region.
The
extended photoionisation front and photon-dominated region seen
edge-on in M17 was modelled by~\citet{PelBalBroEA07} as being
supported by magnetic pressure (observations by~\citealt{BroTro01}
suggest a LOS magnetic field $B>100\,\mu$G).  Unfortunately much of
this H~\textsc{II} region is heavily obscured at optical wavelengths and so
straightforward inspection of $\mathrm{H}\alpha$ images is not particularly
revealing.  In the extensive $\mathrm{H}\alpha$ survey of the Carina nebula
by~\citet*{SmiBalWal10}, only the elephant trunk in `Pos. 23' of
fig.~1 shows evidence of a bright ridge in front of the trunk,
suggesting that such features are not common.  A clearer example is
seen in NGC 6357~\citep{BohTapRotEA04} where a prominent bright ridge
in front of an elephant trunk is the brightest $\mathrm{H}\alpha$ structure in
the field of view.  A similar structure is found in front of a massive
pillar in NGC 3603~\citep{BraGreChuEA00}. In both of these
observations, however, the adjacent massive star cluster is expected
to drive out-flowing gas and so it is unclear if the bright ridges are
ram-pressure or magnetically confined.

The larger simulation domains used here (compared to ML10) allow us to
follow the evolution of the dense gas for significantly longer, up to
$700\,$kyr in some simulations.  We find that in the absence of dense
gas further from the star, a pillar-like structure will be flattened
into a cometary globule with a dense head and low density tail.  To
study this in more detail, clump configuration 2 was set up with an
extra clump a further parsec from the radiation source.  It was found
that the pillar-like stage of the evolution lasted $\sim600\,$kyr in
this model, until all of the dense gas reached the position of the
furthest clump.  Subsequently a cometary morphology developed again.
This experiment shows that while the pillar's lifetime can be extended
somewhat, it cannot survive indefinitely unless there is a long dense
filament pointing away from the radiation source.  This supports the
general picture of elephant trunks shown in fig.~15
of~\citet{SmiPovWhiEA10} as structures which form and disperse over
about $10^6\,$yr.  With our non-gravitating simulations we cannot
model the star formation with also occurs, but the simulations
by~\citet{GriBurNaaEA10} and~\citet{BisWhiWueEA10} show that the compression induced by RDI
produces gravitationally unstable fragments which would likely form
stars.  While it may seem unlikely that the three pillars in M16
should all have originated from elongated overdensities pointing back
towards the ionising stars, recent observations from the Herschel
observatory~\citep[e.g.][]{MolSwiBalEA10} have shown that molecular
clouds appear to have a distinctly filamentary structure.  Simulations
of MHD turbulence generated by colliding flows~\citep{BanVazHenEA09}
show similar structures, and simulations of small-scale non-ideal
isothermal MHD turbulence~\citep{DowOSul09} also show significantly
more linear structure compared to purely hydrodynamical models.

\vspace{-0.5cm}
% ***************************************************
\section{Conclusions}
\label{sec:conclusions}
% ***************************************************
We have performed a series of R-MHD simulations of the photoionisation
of dense clumps of gas and their evolution from pillar-like to
cometary globule-like structures.  Our results for the emissivity of
ionised gas agree very well with those of \citet{HenArtDeCEA09},
showing that a dense, ionised, bar-shaped region standing off from the
ionisation front is a generic feature of strongly magnetised
photoionisation in a clumpy medium for a perpendicular field
orientation.  This ridge can be as bright as, or even brighter than, the
photoionisation front when observed in recombination radiation
(e.g.~$\mathrm{H}\alpha$) and its presence or absence can be used as a
diagnostic of the strength of any large scale magnetic field which may
be present.  Bright ridges or ribbons are observed in some H~\textsc{II}
regions~\citep[e.g.][]{BraGreChuEA00, BohTapRotEA04, SmiBalWal10},
although they are not common.  An overdense ridge could also be produced by
ram-pressure confinement, and more detailed modelling is required to
find observational signatures which could distinguish these different
confinement mechanisms.

Comparing to observations of M16~\citep{HesScoSanEA96} there is no
such ribbon or ridge, suggesting the ambient field measured
by~\citet{SugWatTamEA07} is not dominant in the ionised gas.
This conclusion is strengthened when we consider the magnetic field
orientation observed in M16~\citep[][fig.~9]{SugWatTamEA07}.  The
results presented here show that both RDI and acceleration of clumps
by the rocket effect tend to align the magnetic field in dense neutral
gas with the radiation propagation direction.  In our models a field
configuration similar to the observed one is clearly seen when the
initial field strength is $\vert\mathbf{B}\vert\simeq20\,\mu$G; the
simulation with $\vert\mathbf{B}\vert\simeq50\,\mu$G is consistent with
observations, and the simulation with $\vert\mathbf{B}\vert\simeq160\,\mu$G
is not consistent.  Our simulations thus suggest an ambient field
strength of $\vert\mathbf{B}\vert\lesssim 50\,\mu$G around the M16 pillars.

The morphology of the structures which develop due to RDI and the
rocket effect is also affected by a strong magnetic field, partly due
to shielding by the dense ionised ridge and partly by the effect of the field
within the pillar or globule.  Inspection of $\mathrm{H}\alpha$ images of
elephant trunks and globules in the
literature~\citep[e.g.][]{HesScoSanEA96, SmiBalWal10} suggests that
the features seen in the strong field simulations are not common,
although the uniform initial field configurations considered here are
certainly somewhat artificial.  Additionally many H~\textsc{II} regions have
significantly higher gas pressure than that modelled in our
simulations, in which case the magnetic field must also be
correspondingly stronger to dominate the dynamics.

Comparing these results with our earlier simulations in ML10, the
larger simulation domains used here show that the pillar-like
structures which form will ultimately evolve to cometary structures in
the absence of dense gas further from the star.  The lifetimes of
pillars in our models are $t \lesssim 500\,$kyr, although this depends
significantly on the initial mass and concentration (and presumably
velocity, cf.~\citealt{GriBurNaaEA10}) of the dense gas clumps.

Finally we emphasise, in agreement with previous
authors~\citep{Wil07,KruStoGar07,HenArtDeCEA09}, that a strong
magnetic field has a very significant influence on the dynamics of the
photoionisation process, and many of these effects should be easily
observable.  Given the difficulty of measuring the full 3D magnetic
field in the ISM, comparison to detailed numerical simulations such as
these offers an indirect means to constrain the field strength and
orientation in and around H~\textsc{II} regions.

\vspace{-0.5cm}
% ***************************************************
\section*{Acknowledgments}
% ***************************************************
JM's work has been part funded by the Irish Research Council for
Science, Engineering and Technology; also by a grant from the Dublin
Institute for Advanced Studies, and by Science Foundation Ireland.
AJL's work was funded by a Schr\"{o}dinger Fellowship from the Dublin
Institute for Advanced Studies.  JM acknowledges support from an
Argelander Fellowship during the writing of this paper.  Figures were
generated using the \textsc{VisIt} visualisation tool.  The authors
wish to acknowledge the SFI/HEA Irish Centre for High-End Computing
(ICHEC) for the provision of computational facilities and support.
We thank the referee for useful suggestions and for pointing out an
error in an earlier draft.

\vspace{-0.5cm}
% ***************************************************
\bibliography{refs}

\begin{thebibliography}{}

\bibitem[\protect\citeauthoryear{{Arthur}, {Henney},
  {Mellema}, {De Colle} \& {V{\'a}zquez-Semadeni}}{{Arthur} et~al.}{2010}]{ArtHenMelEA10}
{Arthur} S.~J., {Henney} W.~J., {Mellema} G., {De Colle} F.,
{V{\'a}zquez-Semadeni} E., 2010, \mnras, submitted.

\bibitem[\protect\citeauthoryear{{Banerjee}, {V{\'a}zquez-Semadeni},
  {Hennebelle} \& {Klessen}}{{Banerjee} et~al.}{2009}]{BanVazHenEA09}
{Banerjee} R.,  {V{\'a}zquez-Semadeni} E.,  {Hennebelle} P.,    {Klessen}
  R.~S.,  2009, \mnras, 398, 1082

\bibitem[\protect\citeauthoryear{{Bertoldi}}{{Bertoldi}}{1989}]{Ber89}
{Bertoldi} F.,  1989, \apj, 346, 735

\bibitem[\protect\citeauthoryear{{Bertoldi} \& {McKee}}{{Bertoldi} \&
  {McKee}}{1990}]{BerMcK90}
{Bertoldi} F.,  {McKee} C.~F.,  1990, \apj, 354, 529

\bibitem[\protect\citeauthoryear{{Bhatt}}{{Bhatt}}{1999}]{Bha99}
{Bhatt} H.~C.,  1999, \mnras, 308, 40

\bibitem[\protect\citeauthoryear{{Bhatt}, {Maheswar} \& {Manoj}}{{Bhatt}
  et~al.}{2004}]{BhaMahMan04}
{Bhatt} H.~C.,  {Maheswar} G.,    {Manoj} P.,  2004, \mnras, 348, 83

\bibitem[\protect\citeauthoryear{{Bisbas}, {Whitworth}, {W{\"u}nsch}, {Hubber}
  \& {Walch}}{{Bisbas} et~al.}{2010}]{BisWhiWueEA10}
{Bisbas} T.~G.,  {Whitworth} A.~P.,  {W{\"u}nsch} R.,  {Hubber} D.~A.,
  {Walch} S.,  2010, ArXiv e-prints

\bibitem[\protect\citeauthoryear{{Bodenheimer}, {Tenorio-Tagle} \&
  {Yorke}}{{Bodenheimer} et~al.}{1979}]{BodTenYor79}
{Bodenheimer} P.,  {Tenorio-Tagle} G.,    {Yorke} H.~W.,  1979, \apj, 233, 85

\bibitem[\protect\citeauthoryear{{Bohigas}, {Tapia}, {Roth} \&
  {Ruiz}}{{Bohigas} et~al.}{2004}]{BohTapRotEA04}
{Bohigas} J.,  {Tapia} M.,  {Roth} M.,    {Ruiz} M.~T.,  2004, \aj, 127, 2826

\bibitem[\protect\citeauthoryear{{Brandner}, {Grebel}, {Chu}, {Dottori},
  {Brandl}, {Richling}, {Yorke}, {Points} \& {Zinnecker}}{{Brandner}
  et~al.}{2000}]{BraGreChuEA00}
{Brandner} W.,  {Grebel} E.~K.,  {Chu} Y.,  {Dottori} H.,  {Brandl} B.,
  {Richling} S.,  {Yorke} H.~W.,  {Points} S.~D.,    {Zinnecker} H.,  2000,
  \aj, 119, 292

\bibitem[\protect\citeauthoryear{{Brio} \& {Wu}}{{Brio} \&
  {Wu}}{1988}]{BriWu88}
{Brio} M.,  {Wu} C.~C.,  1988, Journal of Computational Physics, 75, 400

\bibitem[\protect\citeauthoryear{{Brogan} \& {Troland}}{{Brogan} \&
  {Troland}}{2001}]{BroTro01}
{Brogan} C.~L.,  {Troland} T.~H.,  2001, \apj, 560, 821

\bibitem[\protect\citeauthoryear{{Cargo} \& {Gallice}}{{Cargo} \&
  {Gallice}}{1997}]{CarGal97}
{Cargo} P.,  {Gallice} G.,  1997, Journal of Computational Physics, 136, 446

\bibitem[\protect\citeauthoryear{{Carlqvist}, {Gahm} \& {Kristen}}{{Carlqvist}
  et~al.}{2002}]{CarGahKri02}
{Carlqvist} P.,  {Gahm} G.~F.,    {Kristen} H.,  2002, \apss, 280, 405

\bibitem[\protect\citeauthoryear{{Carlqvist}, {Gahm} \& {Kristen}}{{Carlqvist}
  et~al.}{2003}]{CarGahKri03}
{Carlqvist} P.,  {Gahm} G.~F.,    {Kristen} H.,  2003, \aap, 403, 399

\bibitem[\protect\citeauthoryear{{Carlqvist}, {Kristen} \& {Gahm}}{{Carlqvist}
  et~al.}{1998}]{CarKriGah98}
{Carlqvist} P.,  {Kristen} H.,    {Gahm} G.~F.,  1998, \aap, 332, L5

\bibitem[\protect\citeauthoryear{{de Hoffmann} \& {Teller}}{{de Hoffmann} \&
  {Teller}}{1950}]{deHofTel50}
{de Hoffmann} F.,  {Teller} E.,  1950, Physical Review, 80, 692

\bibitem[\protect\citeauthoryear{{Dedner}, {Kemm}, {Kr{\"o}ner}, {Munz},
  {Schnitzer} \& {Wesenberg}}{{Dedner} et~al.}{2002}]{DedKemKroEA02}
{Dedner} A.,  {Kemm} F.,  {Kr{\"o}ner} D.,  {Munz} C.-D.,  {Schnitzer} T.,
  {Wesenberg} M.,  2002, Journal of Computational Physics, 175, 645

\bibitem[\protect\citeauthoryear{{Downes} \& {O'Sullivan}}{{Downes} \&
  {O'Sullivan}}{2009}]{DowOSul09}
{Downes} T.~P.,  {O'Sullivan} S.,  2009, \apj, 701, 1258

\bibitem[\protect\citeauthoryear{{Esquivel} \& {Raga}}{{Esquivel} \&
  {Raga}}{2007}]{EsqRag07}
{Esquivel} A.,  {Raga} A.~C.,  2007, \mnras, 377, 383

\bibitem[\protect\citeauthoryear{{Falle}, {Komissarov} \& {Joarder}}{{Falle}
  et~al.}{1998}]{FalKomJoa98}
{Falle} S.,  {Komissarov} S.,    {Joarder} P.,  1998, \mnras, 297, 265

\bibitem[\protect\citeauthoryear{{Gahm}, {Carlqvist}, {Johansson} \&
  {Nikoli{\'c}}}{{Gahm} et~al.}{2006}]{GahCarJohEA06}
{Gahm} G.~F.,  {Carlqvist} P.,  {Johansson} L.~E.~B.,    {Nikoli{\'c}} S.,
  2006, \aap, 454, 201

\bibitem[\protect\citeauthoryear{{Garcia-Segura} \& {Franco}}{{Garcia-Segura}
  \& {Franco}}{1996}]{GarSegFra96}
{Garcia-Segura} G.,  {Franco} J.,  1996, \apj, 469, 171

\bibitem[\protect\citeauthoryear{{Goodman}, {Jones}, {Lada} \&
  {Myers}}{{Goodman} et~al.}{1995}]{GooJonLadEA95}
{Goodman} A.~A.,  {Jones} T.~J.,  {Lada} E.~A.,    {Myers} P.~C.,  1995, \apj,
  448, 748

\bibitem[\protect\citeauthoryear{{Gritschneder}, {Burkert}, {Naab} \&
  {Walch}}{{Gritschneder} et~al.}{2010}]{GriBurNaaEA10}
{Gritschneder} M.,  {Burkert} A.,  {Naab} T.,    {Walch} S.,  2010, \apj, 723,
  971

\bibitem[\protect\citeauthoryear{{Gritschneder}, {Naab}, {Walch}, {Burkert} \&
  {Heitsch}}{{Gritschneder} et~al.}{2009}]{GriNaaWalEA09}
{Gritschneder} M.,  {Naab} T.,  {Walch} S.,  {Burkert} A.,    {Heitsch} F.,
  2009, \apjl, 694, L26

\bibitem[\protect\citeauthoryear{{Henney}, {Arthur}, {de Colle} \&
  {Mellema}}{{Henney} et~al.}{2009}]{HenArtDeCEA09}
{Henney} W.~J.,  {Arthur} S.~J.,  {de Colle} F.,    {Mellema} G.,  2009,
  \mnras, 398, 157

\bibitem[\protect\citeauthoryear{{Henney}, {Arthur} \&
  {Garc{\'{\i}}a-D{\'{\i}}az}}{{Henney} et~al.}{2005}]{HenArtGar05}
{Henney} W.~J.,  {Arthur} S.~J.,    {Garc{\'{\i}}a-D{\'{\i}}az} M.~T.,  2005,
  \apj, 627, 813

\bibitem[\protect\citeauthoryear{{Hester}, {Scowen}, {Sankrit}, {Lauer},
  {Ajhar}, {Baum}, {Code}, {Currie}, {Danielson}, {Ewald} \& {et al}}{{Hester}
  et~al.}{1996}]{HesScoSanEA96}
{Hester} J.~J.,  et al.,  1996, \aj, 111, 2349

\bibitem[\protect\citeauthoryear{{Hummer}}{{Hummer}}{1994}]{Hum94}
{Hummer} D.~G.,  1994, \mnras, 268, 109

\bibitem[\protect\citeauthoryear{{Indebetouw}, {Robitaille}, {Whitney},
  {Churchwell}, {Babler}, {Meade}, {Watson} \& {Wolfire}}{{Indebetouw}
  et~al.}{2007}]{IndRobWhiEA07}
{Indebetouw} R.,  {Robitaille} T.~P.,  {Whitney} B.~A.,  {Churchwell} E.,
  {Babler} B.,  {Meade} M.,  {Watson} C.,    {Wolfire} M.,  2007, \apj, 666,
  321

\bibitem[\protect\citeauthoryear{{Kessel-Deynet} \& {Burkert}}{{Kessel-Deynet}
  \& {Burkert}}{2003}]{KesBur03}
{Kessel-Deynet} O.,  {Burkert} A.,  2003, \mnras, 338, 545

\bibitem[\protect\citeauthoryear{{Krumholz}, {Stone} \& {Gardiner}}{{Krumholz}
  et~al.}{2007}]{KruStoGar07}
{Krumholz} M.,  {Stone} J.,    {Gardiner} T.,  2007, \apj, 671, 518

\bibitem[\protect\citeauthoryear{{Kuiper}, {Klahr}, {Dullemond}, {Kley} \&
  {Henning}}{{Kuiper} et~al.}{2010}]{KuiKlaDulEA10}
{Kuiper} R.,  {Klahr} H.,  {Dullemond} C.,  {Kley} W.,    {Henning} T.,  2010,
  \aap, 511, A81

\bibitem[\protect\citeauthoryear{{Lasker}}{{Lasker}}{1966a}]{Las66a}
{Lasker} B.~M.,  1966a, \apj, 143, 700

\bibitem[\protect\citeauthoryear{{Lasker}}{{Lasker}}{1966b}]{Las66b}
{Lasker} B.~M.,  1966b, \apj, 146, 471

\bibitem[\protect\citeauthoryear{{Lee} \& {Chen}}{{Lee} \&
  {Chen}}{2007}]{LeeChe07}
{Lee} H.,  {Chen} W.~P.,  2007, \apj, 657, 884

\bibitem[\protect\citeauthoryear{{Lee}, {Chen}, {Zhang} \& {Hu}}{{Lee}
  et~al.}{2005}]{LeeCheZhaEA05}
{Lee} H.,  {Chen} W.~P.,  {Zhang} Z.,    {Hu} J.,  2005, \apj, 624, 808

\bibitem[\protect\citeauthoryear{{Lefloch} \& {Lazareff}}{{Lefloch} \&
  {Lazareff}}{1994}]{LeFLaz94}
{Lefloch} B.,  {Lazareff} B.,  1994, \aap, 289, 559

\bibitem[\protect\citeauthoryear{{Lim} \& {Mellema}}{{Lim} \&
  {Mellema}}{2003}]{LimMel03}
{Lim} A.,  {Mellema} G.,  2003, \aap, 405, 189

\bibitem[\protect\citeauthoryear{{Lora}, {Raga} \& {Esquivel}}{{Lora}
  et~al.}{2009}]{LorRagEsq09}
{Lora} V.,  {Raga} A.~C.,    {Esquivel} A.,  2009, \aap, 503, 477

\bibitem[\protect\citeauthoryear{{Mackey} \& {Lim}}{{Mackey} \&
  {Lim}}{2010}]{MacLim10}
{Mackey} J.,  {Lim} A.~J.,  2010, \mnras, 403, 714

\bibitem[\protect\citeauthoryear{{Marsh}}{{Marsh}}{1970}]{Mar70}
{Marsh} M.~C.,  1970, \mnras, 147, 95

\bibitem[\protect\citeauthoryear{{Mellema}, {Arthur}, {Henney}, {Iliev} \&
  {Shapiro}}{{Mellema} et~al.}{2006b}]{MelArtHenEA06}
{Mellema} G.,  {Arthur} S.,  {Henney} W.,  {Iliev} I.,    {Shapiro} P.,  2006,
  \apj, 647, 397

\bibitem[\protect\citeauthoryear{{Mellema}, {Iliev}, {Alvarez} \&
  {Shapiro}}{{Mellema} et~al.}{2006a}]{MelIliAlvEA06}
{Mellema} G.,  {Iliev} I.,  {Alvarez} M.,    {Shapiro} P.,  2006, New
  Astronomy, 11, 374

\bibitem[\protect\citeauthoryear{{Miao}, {White}, {Nelson}, {Thompson} \&
  {Morgan}}{{Miao} et~al.}{2006}]{MiaWhiNelEA06}
{Miao} J.,  {White} G.~J.,  {Nelson} R.,  {Thompson} M.,    {Morgan} L.,  2006,
  \mnras, 369, 143

\bibitem[\protect\citeauthoryear{{Molinari}, {Swinyard}, {Bally}, {Barlow},
  {Bernard}, {Martin}, {Moore}, {Noriega-Crespo}, {Plume}, {Testi} \& {et
  al}}{{Molinari} et~al.}{2010}]{MolSwiBalEA10}
{Molinari} S.,  et~al.,  2010, \aap, 518, L100

\bibitem[\protect\citeauthoryear{{O'Dell}}{{O'Dell}}{2001}]{ODel01}
{O'Dell} C.~R.,  2001, \araa, 39, 99

\bibitem[\protect\citeauthoryear{{Oort} \& {Spitzer} Jr.}{{Oort} \&
  {Spitzer}}{1955}]{OorSpi55}
{Oort} J.~H.,  {Spitzer} Jr. L.,  1955, \apj, 121, 6

\bibitem[\protect\citeauthoryear{{Orszag} \& {Tang}}{{Orszag} \&
  {Tang}}{1979}]{OrsTan79}
{Orszag} S.~A.,  {Tang} C.,  1979, Journal of Fluid Mechanics, 90, 129

\bibitem[\protect\citeauthoryear{{Osterbrock}}{{Osterbrock}}{1989}]{Ost89}
{Osterbrock} D.~E.,  1989, {Astrophysics of gaseous nebulae and active galactic
  nuclei}.
University Science Books, Mill Valley, CA

\bibitem[\protect\citeauthoryear{{Pellegrini}, {Baldwin}, {Brogan}, {Hanson},
  {Abel}, {Ferland}, {Nemala}, {Shaw} \& {Troland}}{{Pellegrini}
  et~al.}{2007}]{PelBalBroEA07}
{Pellegrini} E.~W.,  et al.,
  2007, \apj, 658, 1119

\bibitem[\protect\citeauthoryear{{Pottasch}}{{Pottasch}}{1958}]{Pot58}
{Pottasch} S.~R.,  1958, \bain, 14, 29

\bibitem[\protect\citeauthoryear{{Pound}}{{Pound}}{1998}]{Pou98}
{Pound} M.~W.,  1998, \apjl, 493, L113

\bibitem[\protect\citeauthoryear{{Pound}, {Kane}, {Ryutov}, {Remington} \&
  {Mizuta}}{{Pound} et~al.}{2007}]{PouKanRyuEA07}
{Pound} M.~W.,  {Kane} J.~O.,  {Ryutov} D.~D.,  {Remington} B.~A.,    {Mizuta}
  A.,  2007, \apss, 307, 187

\bibitem[\protect\citeauthoryear{{Raga}, {Henney}, {Vasconcelos}, {Cerqueira},
  {Esquivel} \& {Rodr{\'{\i}}guez-Gonz{\'a}lez}}{{Raga}
  et~al.}{2009}]{RagHenVasEA09}
{Raga} A.~C.,  {Henney} W.,  {Vasconcelos} J.,  {Cerqueira} A.,  {Esquivel} A.,
     {Rodr{\'{\i}}guez-Gonz{\'a}lez} A.,  2009, \mnras, 392, 964

\bibitem[\protect\citeauthoryear{{Redman}, {Williams}, {Dyson}, {Hartquist} \&
  {Fernandez}}{{Redman} et~al.}{1998}]{RedWilDysEA98}
{Redman} M.~P.,  {Williams} R.~J.~R.,  {Dyson} J.~E.,  {Hartquist} T.~W.,
  {Fernandez} B.~R.,  1998, \aap, 331, 1099

\bibitem[\protect\citeauthoryear{{Ritzerveld}}{{Ritzerveld}}{2005}]{Rit05}
{Ritzerveld} J.,  2005, \aap, 439, L23

\bibitem[\protect\citeauthoryear{{Sandford} II, {Whitaker} \&
  {Klein}}{{Sandford} et~al.}{1982}]{SanWhiKle82}
{Sandford} II M.~T.,  {Whitaker} R.~W.,    {Klein} R.~I.,  1982, \apj, 260, 183

\bibitem[\protect\citeauthoryear{{Schmidt-Voigt} \& {Koeppen}}{{Schmidt-Voigt}
  \& {Koeppen}}{1987}]{SchKoe87}
{Schmidt-Voigt} M.,  {Koeppen} J.,  1987, \aap, 174, 211

\bibitem[\protect\citeauthoryear{{Smith}, {Bally} \& {Walborn}}{{Smith}
  et~al.}{2010a}]{SmiBalWal10}
{Smith} N.,  {Bally} J.,    {Walborn} N.~R.,  2010a, \mnras, 405, 1153

\bibitem[\protect\citeauthoryear{{Smith} \& {Brooks}}{{Smith} \&
  {Brooks}}{2007}]{SmiBro07}
{Smith} N.,  {Brooks} K.~J.,  2007, \mnras, 379, 1279

\bibitem[\protect\citeauthoryear{{Smith}, {Povich}, {Whitney}, {Churchwell},
  {Babler}, {Meade}, {Bally}, {Gehrz}, {Robitaille} \& {Stassun}}{{Smith}
  et~al.}{2010b}]{SmiPovWhiEA10}
{Smith} N.,  et al.,  2010b, \mnras, 406, 952

\bibitem[\protect\citeauthoryear{{Spitzer}}{{Spitzer}}{1998}]{Spi98}
{Spitzer} L.,  1998, {Physical Processes in the Interstellar Medium}.
Wiley

\bibitem[\protect\citeauthoryear{{Sridharan}, {Bhatt} \&
  {Rajagopal}}{{Sridharan} et~al.}{1996}]{SriBhaRaj96}
{Sridharan} T.~K.,  {Bhatt} H.~C.,    {Rajagopal} J.,  1996, \mnras, 279, 1191

\bibitem[\protect\citeauthoryear{{Stone} \& {Gardiner}}{{Stone} \&
  {Gardiner}}{2009}]{StoGar09}
{Stone} J.,  {Gardiner} T.,  2009, New Astronomy, 14, 139

\bibitem[\protect\citeauthoryear{{Stone}, {Gardiner}, {Teuben}, {Hawley} \&
  {Simon}}{{Stone} et~al.}{2008}]{StoGarTeuEA08}
{Stone} J.,  {Gardiner} T.,  {Teuben} P.,  {Hawley} J.,    {Simon} J.,  2008,
  \apjs, 178, 137

\bibitem[\protect\citeauthoryear{{Sugitani}, {Watanabe}, {Tamura}, {Kandori},
  {Hough}, {Nishiyama}, {Nakajima}, {Kusakabe}, {Hashimoto}, {Nagayama},
  {Nagashima}, {Kato} \& {Fukuda}}{{Sugitani} et~al.}{2007}]{SugWatTamEA07}
{Sugitani} K.,  et al.,  2007, \pasj, 59, 507

\bibitem[\protect\citeauthoryear{{T{\'o}th}}{{T{\'o}th}}{2000}]{Tot00}
{T{\'o}th} G.,  2000, Journal of Computational Physics, 161, 605

\bibitem[\protect\citeauthoryear{{van Albada}, {van Leer} \& {Roberts}
  Jr.}{{van Albada} et~al.}{1982}]{vAlbvLeeRob82}
{van Albada} G.~D.,  {van Leer} B.,    {Roberts} Jr. W.~W.,  1982, \aap, 108,
  76

\bibitem[\protect\citeauthoryear{{Voronov}}{{Voronov}}{1997}]{Vor97}
{Voronov} G.~S.,  1997, Atomic Data and Nuclear Data Tables, 65, 1

\bibitem[\protect\citeauthoryear{{Whalen} \& {Norman}}{{Whalen} \&
  {Norman}}{2008}]{WhaNor08}
{Whalen} D.,  {Norman} M.,  2008, \apj, 672, 287

\bibitem[\protect\citeauthoryear{{White}, {Nelson}, {Holland}, {Robson},
  {Greaves}, {McCaughrean}, {Pilbratt}, {Balser}, {Oka}, {Sakamoto},
  {Hasegawa}, {McCutcheon}, {Matthews}, {Fridlund}, {Tothill}, {Huldtgren} \&
  {Deane}}{{White} et~al.}{1999}]{WhiNelHolEA99}
{White} G.~J.,  et al.,
  1999, \aap, 342, 233

\bibitem[\protect\citeauthoryear{{Williams}, {Ward-Thompson} \&
  {Whitworth}}{{Williams} et~al.}{2001}]{WilWarWhi01}
{Williams} R.,  {Ward-Thompson} D.,    {Whitworth} A.,  2001, \mnras, 327, 788

\bibitem[\protect\citeauthoryear{{Williams}}{{Williams}}{2002}]{Wil02}
{Williams} R.~J.~R.,  2002, \mnras, 331, 693

\bibitem[\protect\citeauthoryear{{Williams}}{{Williams}}{2007}]{Wil07}
{Williams} R.~J.~R.,  2007, \apss, 307, 179

\bibitem[\protect\citeauthoryear{{Williams} \& {Dyson}}{{Williams} \&
  {Dyson}}{2001}]{WilDys01}
{Williams} R.~J.~R.,  {Dyson} J.~E.,  2001, \mnras, 325, 293

\bibitem[\protect\citeauthoryear{{Williams}, {Dyson} \& {Hartquist}}{{Williams}
  et~al.}{2000}]{WilDysHar00}
{Williams} R.~J.~R.,  {Dyson} J.~E.,    {Hartquist} T.~W.,  2000, \mnras, 314,
  315

\bibitem[\protect\citeauthoryear{{Williams} \& {Henney}}{{Williams} \&
  {Henney}}{2009}]{WilHen09}
{Williams} R.~J.~R.,  {Henney} W.~J.,  2009, \mnras, 400, 263

\bibitem[\protect\citeauthoryear{{Yorke}}{{Yorke}}{1986}]{Yor86}
{Yorke} H.~W.,  1986, \araa, 24, 49

\end{thebibliography}
% ***************************************************

\appendix

\section{Tests of adiabatic MHD algorithms}
\label{app:MHDtests}

\begin{figure*}
  \centering 
  \includegraphics[width=0.98\textwidth]{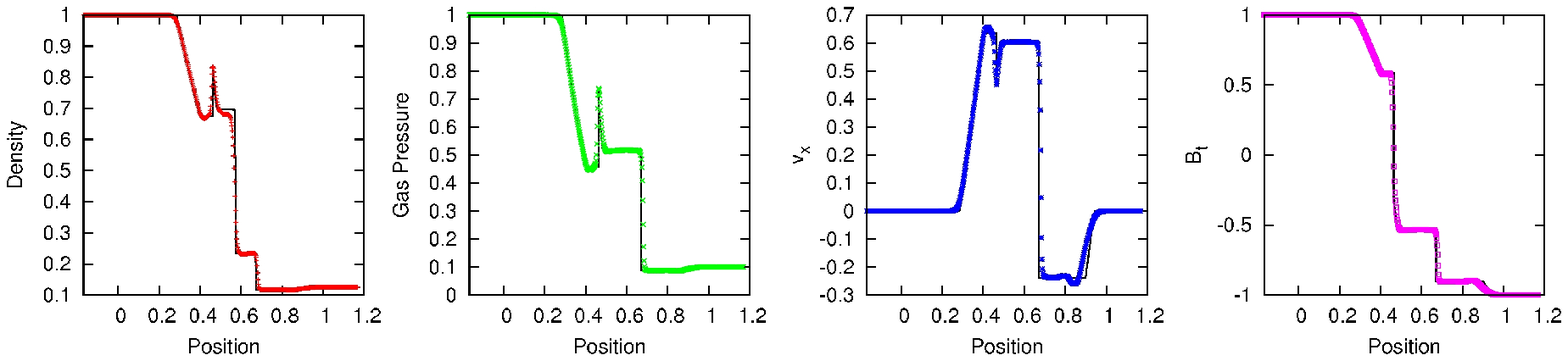}
  \caption{2D shock tube results for the~\citet{BriWu88} adiabatic MHD
    test problem.  The solid line shows the results using 10,000 grid
    cells on a 1D grid, and the points show results in 2D on a domain
    with $200\times200$ cells at time $t=0.12$.  The panels show (from
    left) gas density, pressure, normal velocity and tangential field.
    Note that because all points in the 2D domain are plotted, the
    number of points apparently contained within a discontinuity is
    larger than the actual number of points which resolve it.  }
  \label{fig:BrioWu}
\end{figure*}

\begin{figure*}
  \centering 
  \includegraphics[width=0.45\textwidth]{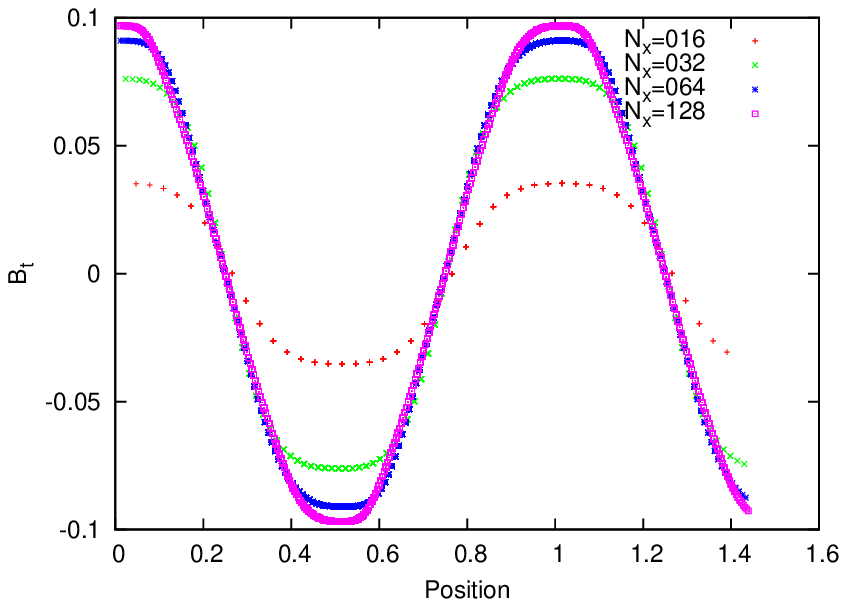}
  \includegraphics[width=0.45\textwidth]{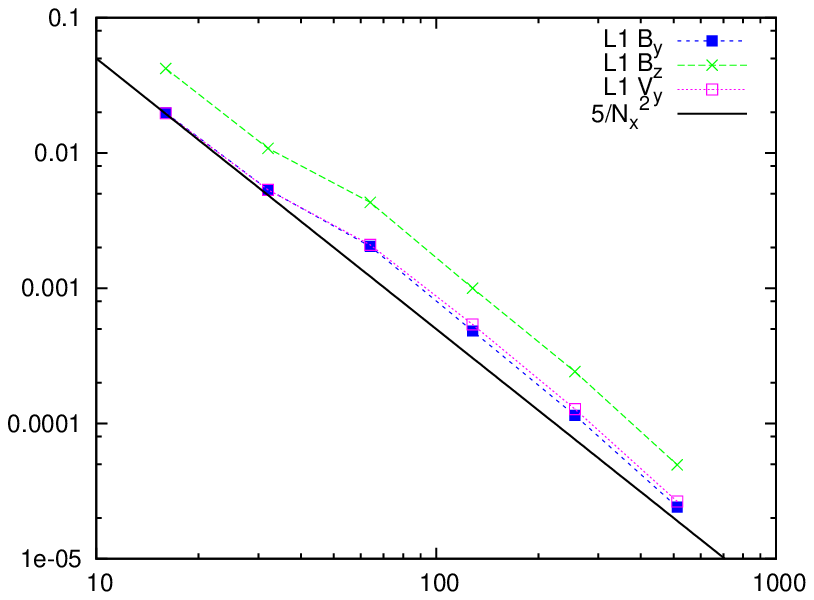}
  \caption{Circularly Polarised Alfv\'en Wave test.  The left-hand
    plot shows the transverse magnetic field after advecting five
    times across the domain (initial amplitude 0.1) for models with
    resolutions $N_x=[16,32,64,128]$.  The right-hand plot shows the
    L1 error (see equation~\ref{eqn:L1error}) as a function of
    resolution for this test, obtained by comparing the initial to the
    final solution at each resolution.  Second order convergence is
    clearly demonstrated.}
  \label{fig:AW2D}
\end{figure*}

Results from the following MHD test problems can be found at
\url{http://www.astro.uni-bonn.de/~jmackey/jmac/} together with the HD
test problems.  These adiabatic calculations are run using
dimensionless units.  The shock-tube tests of~\citet{BriWu88}
and~\citet{FalKomJoa98} were performed in 1D and on a 2D grid with the
shock-normal at an angle $\theta=\tan^{-1}0.5$ to the grid axes,
ensuring that the problem is genuinely multi-dimensional.  2D Results
are shown for the~\citet{BriWu88} test in Fig.~\ref{fig:BrioWu} and
are comparable to those obtained by~\citet{FalKomJoa98}.  The
propagation of a polarised Alfv\'en wave was also calculated in 2D,
using the initial conditions from~\citet{Tot00} and set up on a grid
as in~\citet{StoGarTeuEA08}.  Running this test at different
resolutions shows the degree of numerical diffusion in the algorithm
and also the rate of convergence of the solution.  Results are shown
in Fig.~\ref{fig:AW2D}; comparison of the first panel to results
from~\citet{StoGarTeuEA08} shows that the third order \textsc{Athena}
algorithm is significantly more accurate for this test at a given
resolution, as expected.  The second plot in this figure shows the L1
error:
\begin{equation}
  L_1 =
  \frac{1}{N}\sum_{i=0}^{N-1}\left\vert\phi_i-\phi_i^0\right\vert
\label{eqn:L1error}
\end{equation}
for $N$ cells, where $\phi_i^0$ is the reference state for each cell
and $\phi_i$ is the approximate numerical solution. This verifies that
the code is indeed second order accurate for continuously
differentiable density fields.

Results for the Orszag--Tang vortex problem~\citep{OrsTan79} were also
very similar to those from the \textsc{Athena} code, with slightly
more diffusion for smooth waves because of the lower order of
accuracy.

\begin{figure*}
  \centering 
  \includegraphics[width=0.68\textwidth]{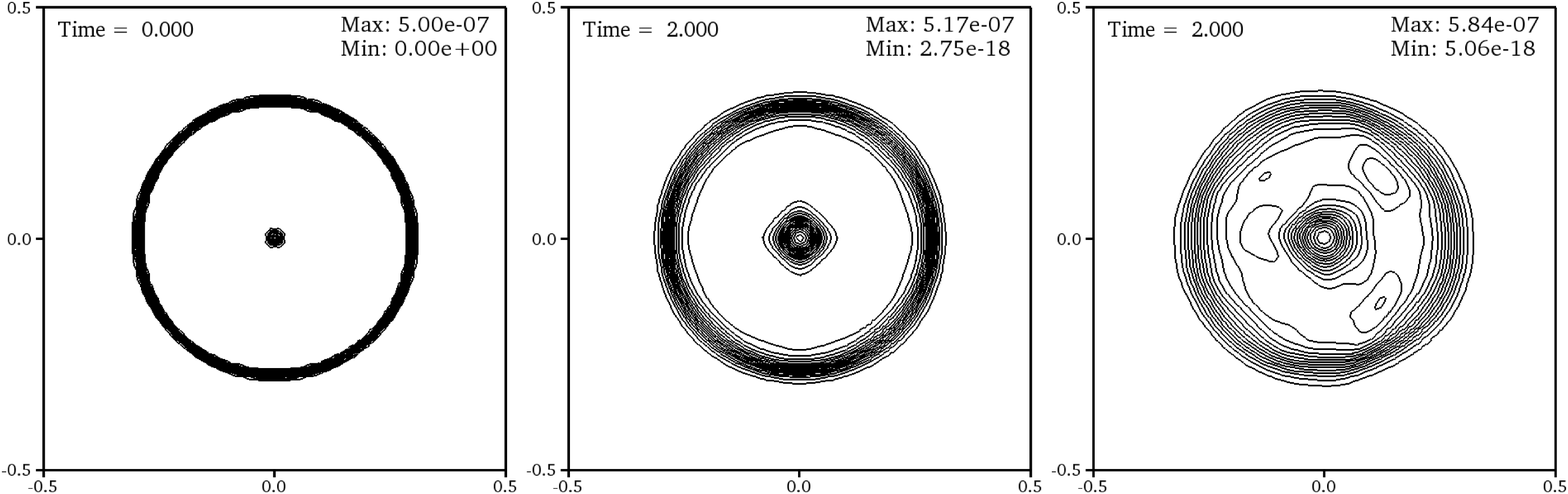}
  \includegraphics[width=0.31\textwidth]{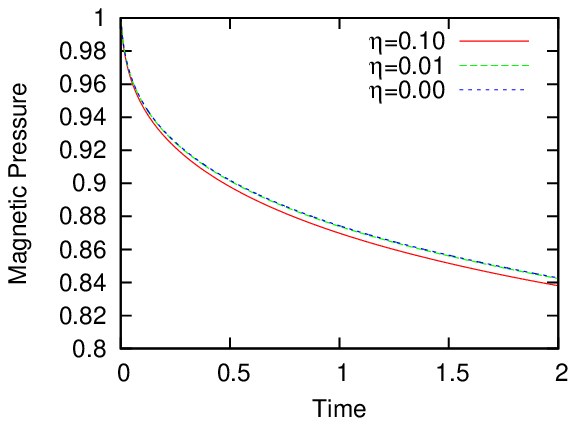}
  \caption{ Field Loop Advection test: the initial (1st panel, from
    left) and final magnetic pressure is shown, with static (2nd
    panel) and advected (3rd panel) models.  Only the central half of
    the domain is shown.  15 contours are shown on a linear scale from
    the minimum to maximum values (dimensionless units).  Some extra diffusion
    and distortion is apparent in the advected model.  The decay of
    magnetic pressure (normalised to the initial value) over time is
    shown in the 4th panel for a model run using the van Albada
    slope-limiter~\citep{vAlbvLeeRob82}.  The three curves show the
    (small) effect of increasing artificial diffusion with the
    viscosity parameter~\citep{FalKomJoa98} indicated.}
  \label{fig:FieldLoop_MagP}
\end{figure*}

The advection of a magnetic field loop~\citep[e.g.][]{StoGarTeuEA08}
across a periodic domain is a challenging test for grid-based codes.
When run with a non-zero velocity in the third dimension ($v_z\neq0$)
this test shows a weakness of the mixed-GLM divergence cleaning method
-- small divergence errors lead to small force errors and some of the
initial magnetic field leaks unphysically into the
$\hat{\mathbf{z}}$-direction ($\sim5$ per cent). Results are shown in the
first three panels of Fig.~\ref{fig:FieldLoop_MagP} for the evolution
of the magnetic pressure.  The left panel shows the initial
conditions; the centre panel shows the state at $t=2$ for a model with
no advection (to show the numerical diffusion), and the right panel
shows the state at $t=2$ for a model advected twice across the
domain. The time-decay of magnetic pressure for the advected models is
shown in the right-most panel of Fig.~\ref{fig:FieldLoop_MagP}. The
constrained transport method employed in the \textsc{Athena} code,
together with its third order accuracy, leads to less diffusion and a
more symmetric solution to this problem than is obtained with our
code~\citep[cf.][]{StoGarTeuEA08}, but our results are comparable to
those of other 2nd order astrophysics MHD codes.  Tests
also showed that for this test the van Albada
slope-limiter~\citep{vAlbvLeeRob82} produced a noticeably superior
solution than the commonly used MinMod limiter, probably because the
latter is significantly more diffusive.

\begin{figure}
  \centering 
 \includegraphics[width=0.23\textwidth]{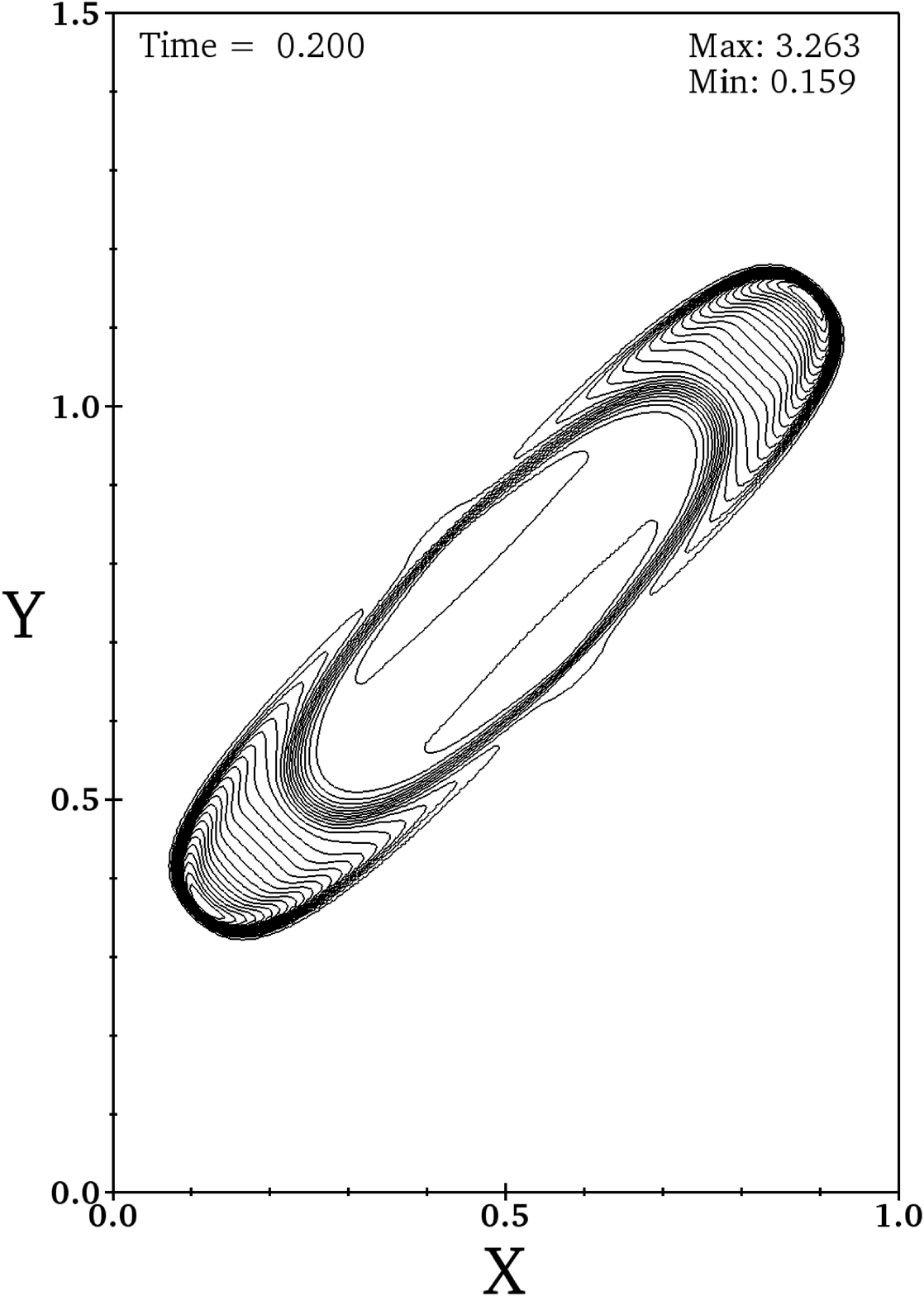}
  \includegraphics[width=0.23\textwidth]{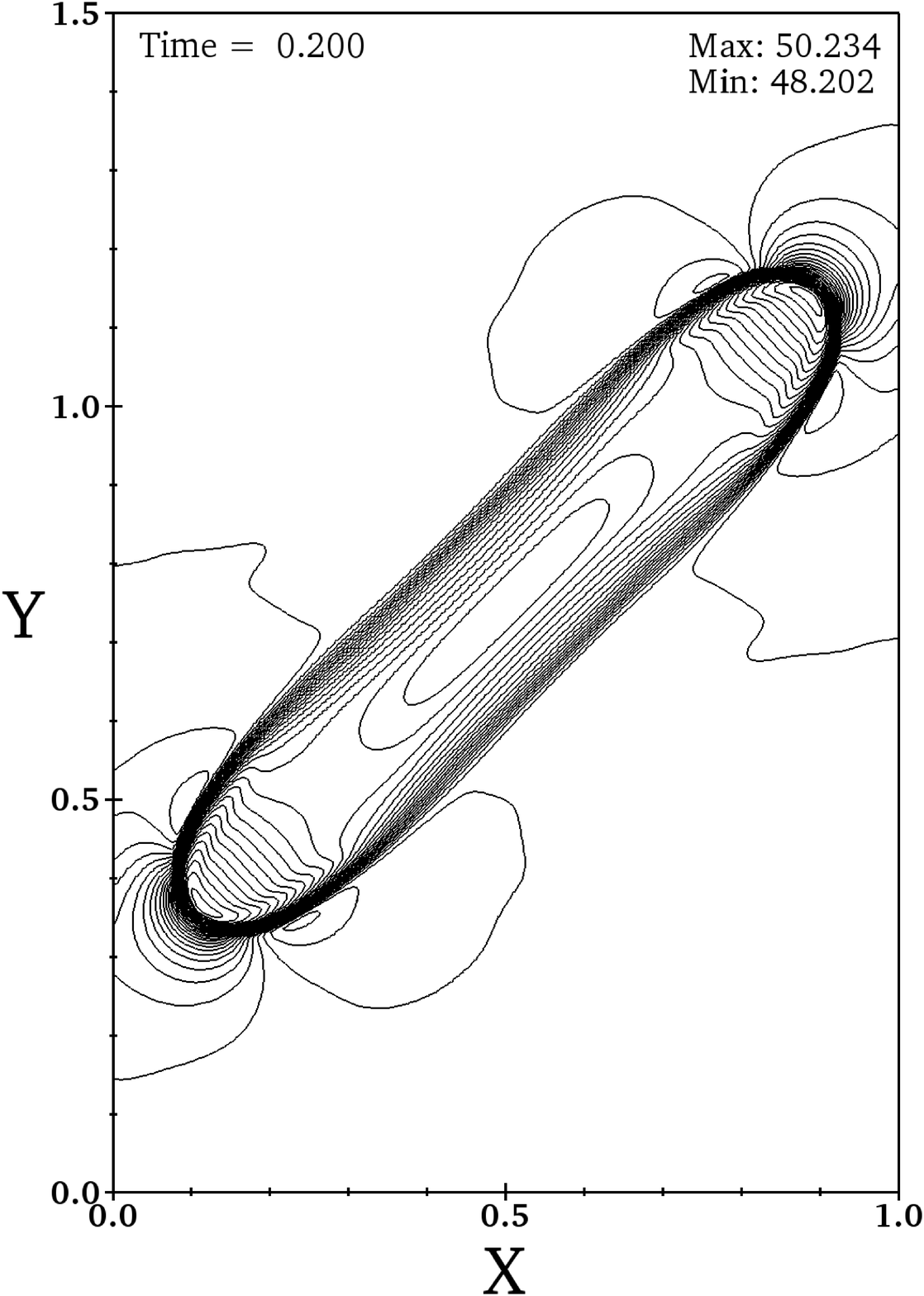}
  \caption{Contour plots of slices through the gas density (left) and
    magnetic pressure (right) at time $t=0.2$ for the MHD blast wave
    test problem with $\beta_0=0.002$.  The slice is through the
    mid-plane $z=0.5$.  There are 30 contours on a linear scale from
    the indicated minimum to maximum values.  Because of the higher
    wavespeeds here compared to the $\beta_0=0.2$ model
    in~\citet{StoGar09}, the fast magnetosonic waves seen in their
    results have long since left the domain in this simulation.}
  \label{fig:SG09BW_Beta0p002}
\end{figure}
 
The expansion of a multi-dimensional adiabatic blast wave provides a
good test of a code's robustness in modelling 3D shocks and
rarefactions. We ran the same 3D problem as described
in~\citet[][\S6.5]{StoGar09}, with the gas pressure in a sphere at
centre of the domain set to $100\times$ the ambient pressure, and with
a uniform magnetic field at $45^{\circ}$ to the grid axes. Very similar
results were found for the case where the initial plasma beta
parameter $\beta=0.2$.  Because of the robustness of
the~\citet{DedKemKroEA02} algorithm, and presumably also because the
integration algorithm used is slightly more diffusive than the
\textsc{Athena} algorithm, our code can also simulate this test
problem with a field $10\times$ stronger ($\beta=0.002$), shown in
Fig.~\ref{fig:SG09BW_Beta0p002}. This extra robustness is important
for the strong field simulations run here.

\section{Boundary condition for mixed-GLM divergence cleaning
  algorithm}
\label{app:PsiBC}
\citet{DedKemKroEA02} suggest that for zero-gradient boundary
conditions (BCs), using zero-gradient for the unphysical scalar field
$\psi$ is sufficient. We have found, however, that more care is needed
for magnetically dominated regions (plasma parameter $\beta\equiv8\pi
p_g/B^2\ll1$) when gas is flowing subsonically near the boundary. In
what follows $\hat{\mathbf{x}}$ is the normal vector to the boundary, the
computational domain is `left' of the boundary, and off the domain is
`right'. Note that $\psi$ is not a physical variable so its gradient
is not constrained. In fact if the field is constant across the
boundary then the formal solution to the 1D Riemann problem is that
the flux, $F[B_x]=0$ (as it is for all 1D Riemann problems). So
ideally a boundary condition for $\psi$ should be selected which
ensures $F[B_x]=0$.  From~\citet[][eqs.~41,42]{DedKemKroEA02} it is
easy to calculate this: the zero-gradient condition states that
$B_x^r-B_x^l=0$ so if additionally $\psi^r=-\psi^l$ then $F[B_x]=0$ is
obtained.

\begin{figure}
  \centering 
  \includegraphics[width=0.4\textwidth]{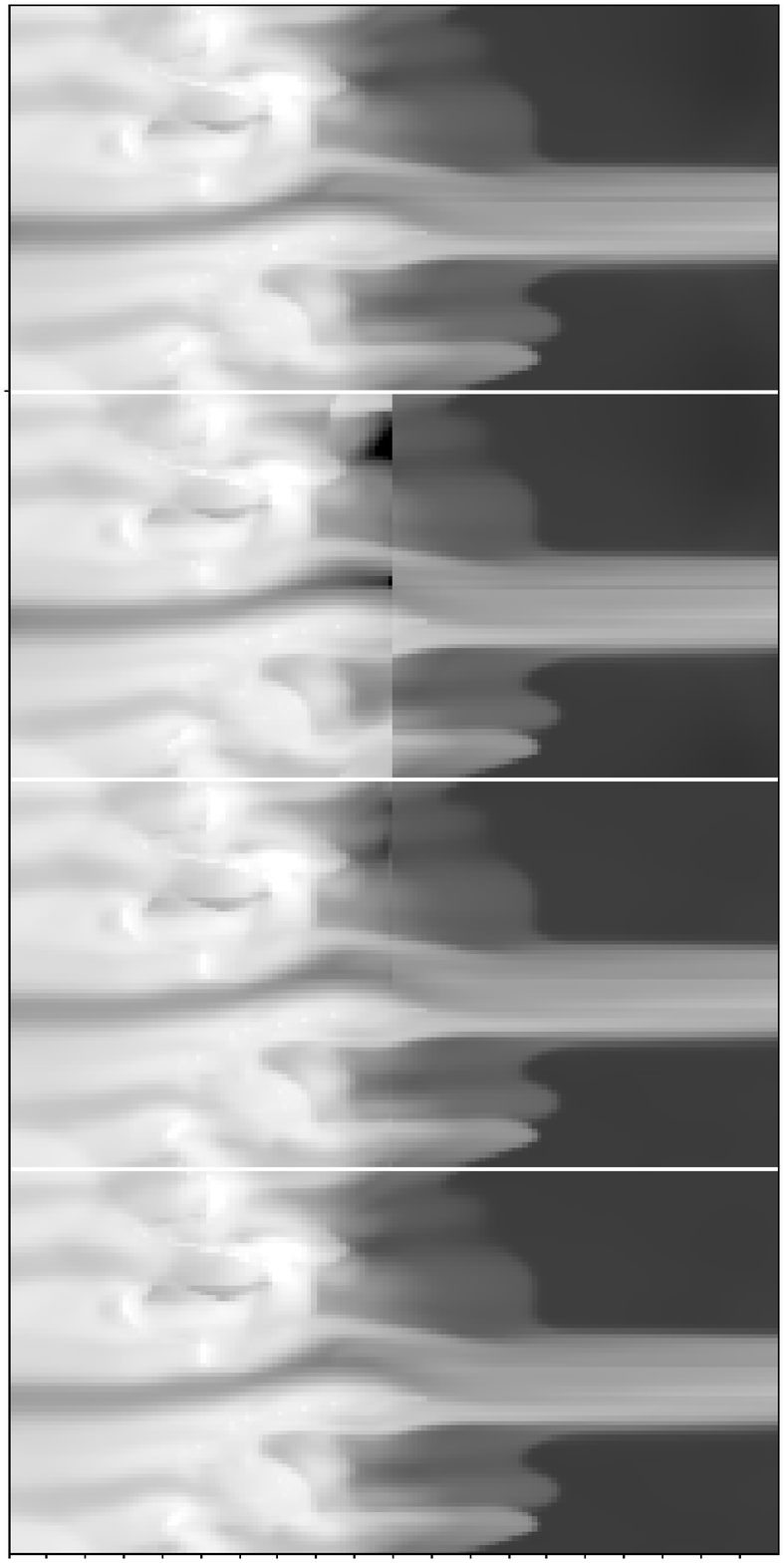}
  \caption{Plots of gas pressure on a log scale from
    $10^{-12}$--$10^{-8}\,\mathrm{dyne}\,.\mathrm{cm}^{-2}$, at time
    $t=12\,$kyr.  Ionised gas has typical pressures of
    $10^{-8}\,\mathrm{dyne}\,.\mathrm{cm}^{-2}$ and neutral gas
    $10^{-11}\,\mathrm{dyne}\,.\mathrm{cm}^{-2}$.  The top panel shows
    the full domain with $\psi^r=\psi^l$ as the BC.  The second panel
    shows the same, but with a simulation of only the left half of the
    domain over-plotted to highlight the boundary effects, again with
    $\psi^r=\psi^l$.  The third panel is as the second, but using
    $\psi^r=-\psi^l$ as the BC for both simulations, and the bottom
    panel shows the full domain with the $\psi^r=-\psi^l$ BC.  The
    simulations were run using cooling model C2.}
  \label{fig:TestBC_PG_c15}
\end{figure}
 
A sequence of figures from a 2D slab-symmetric test simulation (this
time using c.g.s.\ units) are shown in Fig.~\ref{fig:TestBC_PG_c15} to
demonstrate the effectiveness of the new boundary condition. The
simulation domains used have sizes $[0.2,0.1]\,$pc (large) and
$[0.1,0.1]\,$pc (small) and the radiation source is at infinity in the
$-\hat{\mathbf{x}}$ direction.  A group of randomly located dense clumps
are placed between $x=0.02\,$pc and $x=0.08\,$pc on a uniform
background density of $n_{\mathrm{H}}=100\,\mathrm{cm}^{-3}$. The initial magnetic
field is $\mathbf{B} = [50,1,0]\times \sqrt{4\pi}\,\mu$G, and the initially
constant gas pressure is set to give $\beta=0.017$ ($T=1500$K in the
lowest density gas). Zero gradient BCs are enforced in the
$\pm\hat{\mathbf{x}}$ directions, and periodic in $\pm\hat{\mathbf{y}}$.

Fig.~\ref{fig:TestBC_PG_c15} shows the gas pressure at time
$t=12\,$kyr using cooling model C2. The problem with the zero-gradient
BC for $\psi$ is very apparent, as is the dramatic improvement on
switching to the $\psi^r=-\psi^l$ BC. The black regions are hiding
cells which obtained negative pressures and were set to an artificial
pressure floor value in order to allow the simulation to continue.
Regarding the other boundaries, the $x=0.2\,$pc boundary only has
waves moving perpendicular to it and so there is negligible net flow
across the boundary.  The $x=0\,$pc boundary has strong supersonic
outflows, so the details of the BC for $\psi$ have little effect on
it.

\section{Boundary effects as a function of domain size}
\label{app:DomainSize}
\begin{table}
  \begin{tabular}{ l  l  l  l  l  l }
    \hline
    Name & $x_{\mathrm{min}}$ & $x_{\mathrm{max}}$ & $y_{\mathrm{min}}$/$z_{\mathrm{min}}$ &
    $y_{\mathrm{max}}$/$z_{\mathrm{max}}$ & cells \\
    \hline
    Small  & 2.0 & 4.0 & -0.75 & 0.75 & $128\times96^2$ \\
    Medium & 1.5 & 4.5 & -1.125 & 1.125 & $192\times144^2$ \\
    Large  & 1.5 & 5.5 & -1.5 & 1.5 & $256\times192^2$ \\
    X-Long & 1.5 & 6.5 & -1.5 & 1.5 & $320\times192^2$\\
    X-Wide & 1.5 & 5.5 & -2.0 & 2.0 & $256^3$\\
    \hline
  \end{tabular}
  \caption{List of simulations run to test the boundary effects on the
    results of 3D MHD simulations of photoionisation.  All tests were
    run with a $15\times\sqrt{4\pi}\simeq53\,\mu$G magnetic field in the $y$--$z$
    plane.  Boundary positions are measured in parsecs in coordinates
    where the source is at $[0,0,0]$, as in Table~\ref{tab:clump_props}.}
  \label{tab:M17mhd_BoundaryTest} 
\end{table}

\begin{figure}
  \centering 
  \includegraphics[width=0.4\textwidth]{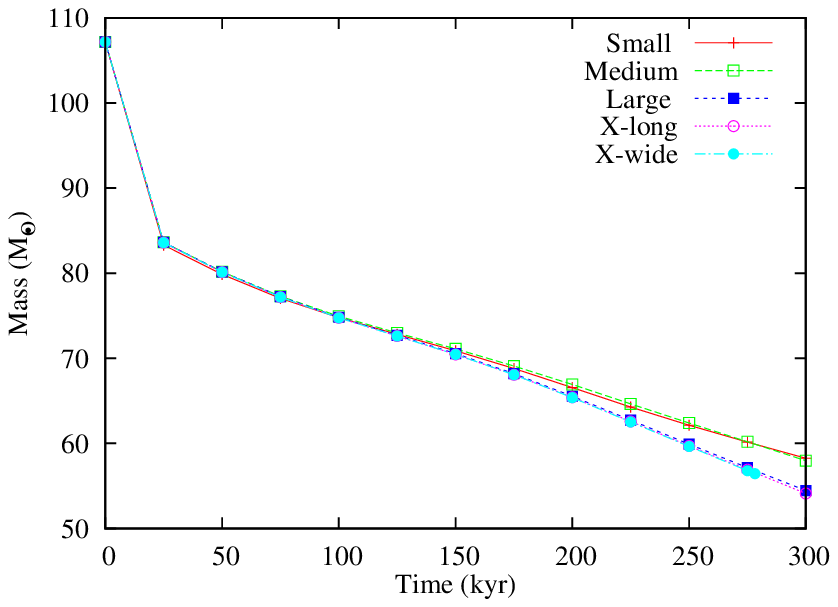} \\
  \includegraphics[width=0.4\textwidth]{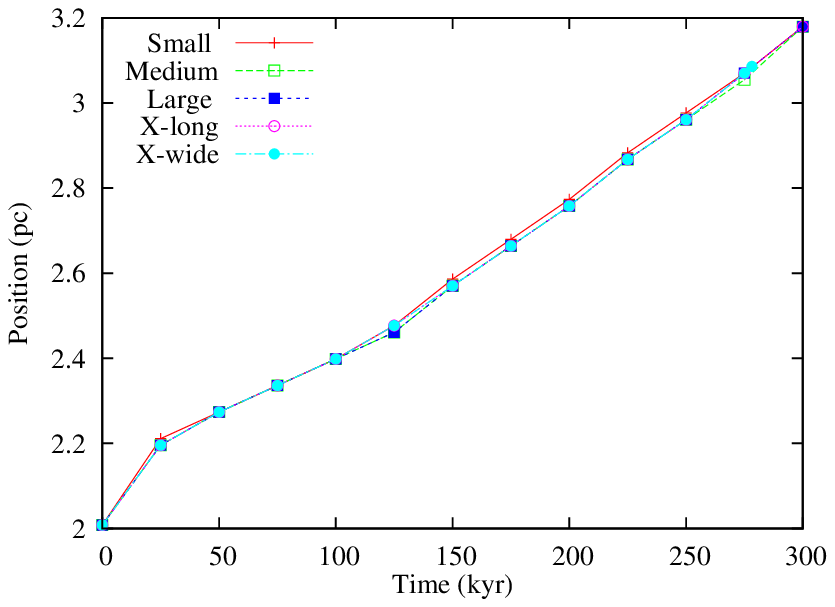} \\
  \includegraphics[width=0.4\textwidth]{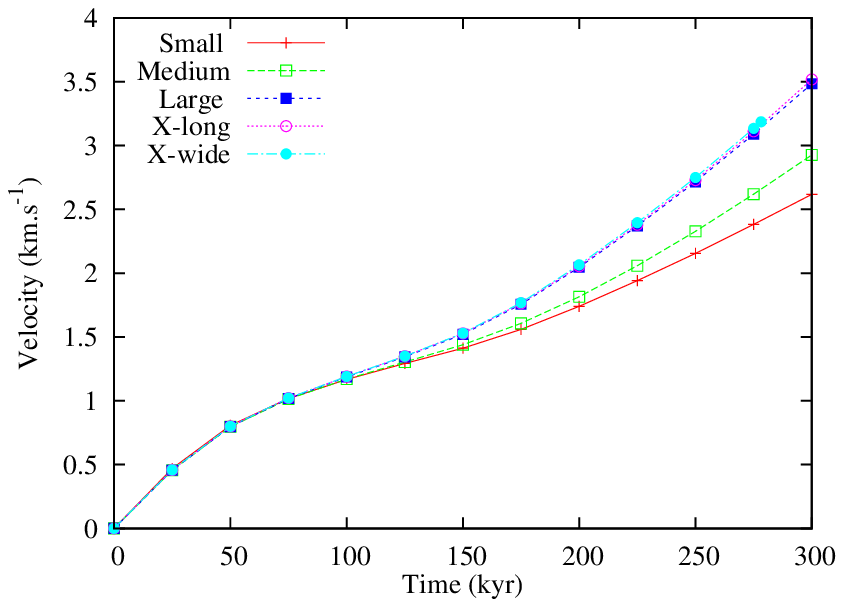} \\
  \caption{(1) Neutral mass, (2) Position of leading edge of
    ionisation front, and (3) mass-weighted mean neutral gas
    $x$-velocity.  Note that the large, extra-long and extra-wide
    simulations have almost converged for all plots.}
  \label{fig:MHDboundary_comp}
\end{figure}
To study boundary effects on the results presented here, test
simulations using clump configuration 1 were performed with
increasingly larger domains, both in 2D and 3D with a medium
perpendicular magnetic field (model R5 in
Table~\ref{tab:rmhd_Fields}).  The same physical resolution was used
for each simulation; boundary positions for five 3D simulations are
listed in Table~\ref{tab:M17mhd_BoundaryTest}.  Various global
quantities were measured within the volume common to all simulations.
Neutral mass, mean neutral gas velocity, and position of the first
neutral cell on the domain are plotted as a function of time in
Fig.~\ref{fig:MHDboundary_comp}.  The small and medium simulations are
clearly significantly affected by the boundaries whereas the large
model is essentially identical to the X-long and X-wide models,
indicating that the results have converged at least in this limited
subdomain of the simulation.  The position of the leading edge of the
ionisation front is unaffected by the position of the boundaries, but
the neutral gas mass and velocity are significantly affected.  It was
found that flow of gas through the distant boundary was impeded in
these R-MHD simulations, something which did not happen in R-HD
simulations.

\end{document}